\documentclass{pasj00}

\Received{$\langle$ $\rangle$}
\Accepted{$\langle$ $\rangle$}
\Published{$\langle$ $\rangle$}
\SetRunningHead{Ishiyama et al.}{The $\nu^2$GC Simulations}

\newcommand{\makiya}{(Makiya et al., in prep)}
\newcommand{\nugc}{$\nu^2$GC}
\newcommand{\ngc}{$\nu$GC}

\begin{document}

\title{The {\boldmath $\nu^2$GC} Simulations : 
Quantifying the Dark Side of the Universe
in the Planck Cosmology}

\author{Tomoaki \textsc{Ishiyama}\altaffilmark{1,2}}
\altaffiltext{1}{Institute of Management and Information Technologies, Chiba University, 
1-33, Yayoi-cho, Inage-ku, Chiba, 263-8522, Japan}
\altaffiltext{2}{Center for Computational Science, University of Tsukuba, 1-1-1,
  Tennodai, Tsukuba, Ibaraki, 305-8577, Japan}
\email{ishiyama@chiba-u.jp}

\author{Motohiro \textsc{Enoki}\altaffilmark{3}}
\altaffiltext{3}{Faculty of Business Administration, Tokyo Keizai University, 
Kokubunji, Tokyo, 185-8502, Japan}

\author{Masakazu A.R. \textsc{Kobayashi}\altaffilmark{4}}
\altaffiltext{4}{
Research Center for Space and Cosmic Evolution, Ehime University, Matsuyama, Ehime, 790-8577, Japan}

\author{Ryu \textsc{Makiya}\altaffilmark{5}}
\altaffiltext{5}{
Institute of Astronomy, the University of Tokyo, Mitaka, Tokyo, 181-0015, Japan
}

\author{Masahiro \textsc{Nagashima}\altaffilmark{6,7}}
\author{Taira \textsc{Oogi}\altaffilmark{6,7}}
\altaffiltext{6}{
Faculty of Education, Nagasaki University, Nagasaki, Nagasaki, 852-8521, Japan
}
\altaffiltext{7}{
Faculty of Education, Bunkyo University, Koshigaya, Saitama 343-8511, Japan}


\maketitle

\begin{abstract} 

We present the evolution of dark matter halos in six large
cosmological {\it N}-body simulations, called the $\nu^2$GC (New
Numerical Galaxy Catalog) simulations on the basis of the $\Lambda$CDM
cosmology consistent with observational results obtained by the Planck
satellite. The largest simulation consists of $8192^3$ (550 billion)
dark matter particles in a box of $1.12 \, h^{-1} \rm Gpc$ (a mass resolution of
$2.20 \times 10^{8} \, h^{-1} M_{\odot}$). Among simulations utilizing
boxes larger than $1 \, h^{-1} \rm Gpc$, our simulation yields the highest
resolution simulation that has ever been achieved. A $\nu^2$GC simulation
with the smallest box consists of eight billions particles in a box of
$70 \, h^{-1} \rm Mpc$ (a mass resolution of $3.44 \times 10^{6} \, h^{-1}
M_{\odot}$). These simulations can follow the evolution of halos over
masses of eight orders of magnitude, from small dwarf galaxies to
massive clusters.  Using the unprecedentedly high resolution and
powerful statistics of the $\nu^2$GC simulations, we provide 
statistical results of the halo mass function, mass accretion rate,
formation redshift, and merger statistics, and present accurate fitting
functions for the Planck cosmology.  
By combining the
$\nu^2$GC simulations with our new semi-analytic galaxy formation model,
we are able to prepare mock catalogs of galaxies and active galactic
nuclei, which will be made publicly available in the near future.

\end{abstract}

\section{Introduction}\label{sec:intro}

Galaxy formation is one of the fundamental processes driving the 
evolution of our Universe. However, the details of its underlying 
mechanisms are not well understood. There are many physical processes 
involved in galaxy formation and evolution such as star formation, 
active galactic 
nuclei (AGN) formation, their feedback, and supernovae.
For example, observations have demonstrated 
that masses of supermassive blackholes in galactic centers correlate 
with the stellar masses and stellar velocity dispersions of galactic 
bulges (e.g., \cite{Magorrian1998}).  This correlation suggests that 
galaxies and supermassive blackholes in galactic centers co-evolve while 
interacting with each other. However, the details of the physical 
processes involved in their interactions are not well understood.

Some cosmological hydrodynamic simulations for galaxy and AGN formation 
have partially succeeded in reproducing observational results such as 
the AGN luminosity function and AGN number density evolution 
(e.g.,\cite{Degraf2010, Hirschmann2014, Khandai2014}).  However, 
sufficient data to demonstrate the result's statistical significance 
are lacking. To improve the mass resolution of these simulations, 
relatively small simulation boxes have been used (less than $100 \, h^{-1} \rm Mpc$), 
making it difficult to follow the formation and evolution of numerous 
galaxies and AGNs at high redshifts.  Since cosmological hydrodynamic 
simulations with large boxes and sufficient mass resolution remain 
computationally challenging, a large gap exists between numerical 
simulations and observations for high redshifts.

One way to bridge this gap is to use a semi-analytic galaxy formation 
model in which galaxy and AGN formation is modeled phenomenologically 
within merger histories of dark matter halos (merger trees) taken from 
cosmological {\it N}-body simulations or the analytic extended 
Press--Schechter model (EPS; \cite{Peacock1990, Bond1991, Bower1991}). 
Examples of successful 
semi-analytic galaxy formation models are the \ngc\ \citep{Nagashima2005} 
model based on the Mitaka model \citep{Nagashima2004} and its successor, 
the \nugc\ model \makiya.  Most recent semi-analytic models use merger 
trees taken from large cosmological {\it N}-body simulations, which can 
easily provide spatial information about galaxies and AGNs.  Generating 
merger trees using the EPS method is appealing because the computational 
cost is low. However, this method does not provide spatial 
information, and its merger statistics are significantly different from 
those of cosmological simulations (e.g., \cite{Jiang2014}).

The rarity of the objects that can be obtained in semi-analytic models
depends on the box size of the adopted cosmological {\it N}-body
simulations. In particular, bright AGNs at high redshifts are rare
objects. Their number density is only $\sim 10^{-8}\mbox{--}10^{-6} \, \rm
{Mpc^{-3} \, mag^{-1}}$ (e.g., \cite{Fontanot2007, Croom2009, Ikeda2011,
  Ikeda2012}).  To obtain a statistically significant number of mock
galaxies and AGNs at high redshifts, a high mass resolution and large
spatial volume are necessary. Our previous study, which was based on a
cosmological simulation with $2048^3$ particles in a $280 \, h^{-1} \rm Mpc$ box,
shows that the AGN downsizing trend naturally emerges
\citep{Enoki2014}.  However, the number of AGNs at $z=2$ was only $\rm
\sim 10^{-6} \, Mpc^{-3} \, mag^{-1} $, complicating the discussion of the
spatial clustering of AGNs and their evolution.  Galaxies at high
redshifts are smaller than those at low redshifts. A high mass
resolution is required to follow the hierarchical formation of these
small galaxies. To compare these results with forthcoming wide and
deep observations of galaxies and AGNs provided by the Subaru
Hyper--Suprime--Cam \citep{Miyazaki2006, Miyazaki2012}, we need
extremely large simulations based on state-of-the-art cosmology.

Most of the simulations in the existing literature do not meet 
these
requirements. One typical simulation suite that succeeded in many
scientific aspects is the Millennium runs \citep{Springel2005, Boylan2009,
  Angulo2012}. The box sizes of the Millennium and Millennium-XXL 
simulations are $500 \, h^{-1} \rm Mpc$ and $3 \, h^{-1} \rm Gpc$, respectively, 
which are large enough to obtain 
a large number of bright AGNs.  However, their mass 
resolutions are too poor to capture small galaxy formations.  The box 
size of the Millennium-II simulation ($100 \, h^{-1} \rm Mpc$) is too small to obtain 
a sufficient number of AGNs, but its mass resolution is better than the 
other Millennium variations.  
Thus, it may be 
difficult to compare mock catalogs of these simulations with forthcoming 
wide and deep AGN observations.
The cosmology of the Millennium runs is 
based on the first year Wilkinson microwave anisotropy probe (WMAP) data 
\citep{Spergel2003}, which differs significantly from recent results 
obtained by the Planck satellite \citep{Planck2014}.  
However, the different cosmology may not be a problem 
by using a rescaling algorithm \citep{Angulo2010, Angulo2014}, 
as also shown in \citet{Guo2013} and \citet{Henriques2014}.

To address these problems, we conducted a suite of ultralarge
cosmological {\it N}-body simulations, the \nugc\ simulation suite.  The
largest volume used in this suite 
is $1.12 \, h^{-1} \rm Gpc$, and the highest mass resolution is $3.44
\times 10^{6} \, h^{-1} M_{\odot}$.  The largest \nugc\ simulation includes
$8192^3=549,755,813,888$ dark matter particles in a box of
$1.12 \, h^{-1} \rm Gpc$. The mass of each particle is $2.20 \times 10^{8}
\, h^{-1} M_{\odot}$.  
Compared with the Millennium simulation
\citep{Springel2005}, 
our simulation offers the advantages of a mass resolution that is four times better 
and a spatial volume that is 11 times larger.  
Simulations utilizing boxes greater than $1 \, h^{-1} \rm Gpc$ have previously achieved 
mass resolutions peaking at 
an order $10^9 \, h^{-1} M_{\odot}$ \citep{Teyssier2009, Angulo2012, Skillman2014}.
Our mass resolution is more than 20 times higher
than other large volume simulations. The cosmology of our simulation is based on 
state-of-the-art observational results obtained by the Planck satellite
\citep{Planck2014}, which provide significantly different
cosmological parameters from those adopted in previous studies.  By
accomplishing unprecedentedly high mass resolution and statistical power, our
simulation is more advanced than any simulation to date and provides the most 
accurate tools for studying galaxy and AGN formation and cosmology.

In this study, we present details of the \nugc\ simulations. In \S
\ref{sec:method}, we describe the basic properties of the simulations and
the applied numerical methods. The basic results, including mass function, mass
accretion rate, statistics of formation redshift, and statistics of
merger, are presented in \S \ref{sec:result}. The results are summarized
in \S \ref{sec:summary}. The details of our \nugc\ semi-analytic model
are also presented in a companion study \makiya.

\section{The \nugc\ Simulation Suite}\label{sec:method}

\subsection{Simulation Details}

The \nugc\ simulation suite consists of six large cosmological
$N$-body simulations with varying mass resolutions and box sizes. 
The details of the six simulations are outlined in Table
\ref{tab:simulations}.
In the largest run, named \nugc-L, we simulated
the motions of $8192^3=549,755,813,888$ dark matter particles with a
mass resolution of $2.20\times 10^8 \, h^{-1} M_{\odot}$ in a comoving box of
$1120 \, h^{-1} \rm Mpc$. We performed two simulations with smaller boxes and the
same mass resolution, namely, $4096^3$ particles in a $560 \, h^{-1} \rm Mpc$ box
(\nugc-M) and $2048^3$ particles in a $280 \, h^{-1} \rm Mpc$ box (\nugc-S). These
smaller simulations allow us to perform resolution studies. In two 
additional runs, we simulated the motions of $2048^3$ particles using
higher mass resolutions and smaller boxes than those of the other three
simulations. We named these simulations \nugc-H1 and \nugc-H2.  The
box size and mass resolution of the former are $140 \, h^{-1} \rm Mpc$ and $2.75
\times 10^7 \, h^{-1} M_{\odot}$, and those of the latter are $70 \, h^{-1} \rm Mpc$ and
$3.44 \times 10^6 \, h^{-1} M_{\odot}$. These five simulations were
terminated at $z=0$.  In the final simulation, named \nugc-H3, we
simulated $4096^3$ particles in a $140 \, h^{-1} \rm Mpc$ box down to $z=4$.  
Since this simulation adopts an equivalent mass resolution 
and eight times larger spatial volume compared with 
those of \nugc-H2
, galaxies may be studied at high
redshifts with unprecedentedly high mass resolutions and statistics.

The initial conditions used in these simulations were generated by a publicly available code,
2LPTic\footnote{http://cosmo.nyu.edu/roman/2LPT/\newline We slightly
  modified the code to enable a more rapid generation of initial conditions with
  over $2048^3$ particles.}, using second-order Lagrangian
perturbation theory (e.g., \cite{Crocce2006}).  The adopted cosmological
parameters were based on an observation of the cosmic 
microwave background obtained by the Planck satellite
\citep{Planck2014}, namely, $\Omega_0=0.31$, $\Omega_b=0.048$,
$\lambda_0=0.69$, $h=0.68$, $n_s=0.96$, and $\sigma_8=0.83$.
To calculate the transfer function, we used the online version\footnote
{http://lambda.gsfc.nasa.gov/toolbox/tb\_camb\_form.cfm}
of CAMB \citep{Lewis2000}.
All simulations began at $z=127$.

\begin{table*}[t]
\centering
\caption
{Details of the \nugc\ simulations. $N$ is the total number of particles, 
$L$ is the comoving box size, $m$ is the particle mass, 
$\varepsilon$ is the Plummer softening length, and 
$M_{\rm min}$ is the mass of the smallest halos identified 
by the FoF algorithm with a linking length of $b=0.2$.
The number of particles for the smallest halos is 40. 
With the exception of the \nugc-H3 simulation, which was terminated at $z=4$, 
all other simulations were terminated at $z=0$. 
}\label{tab:simulations}
\begin{tabular}{lccccc}
\hline
Name & $N$ & $L(h^{-1} \rm Mpc)$ & $m (h^{-1} M_{\odot})$ & $\varepsilon (h^{-1} \rm kpc)$ & $M_{\rm min} (h^{-1} M_{\odot})$ \\
\hline
\nugc-L  & $8192^3=549,755,813,888$ & $1120.0$ & $2.20 \times 10^{8}$ & $4.27$ & $8.79 \times 10^{9}$ \\
\nugc-M  & $4096^3=68,719,476,736$ & $560.0$  & $2.20 \times 10^{8}$ & $4.27$ & $8.79 \times 10^{9}$ \\
\nugc-S  & $2048^3=8,589,934,592$ & $280.0$  & $2.20 \times 10^{8}$ & $4.27$ & $8.79 \times 10^{9}$ \\
\nugc-H1 & $2048^3=8,589,934,592$ & $140.0$  & $2.75 \times 10^{7}$ & $2.14$ & $1.10 \times 10^{9}$ \\
\nugc-H2 & $2048^3=8,589,934,592$ & $70.0$   & $3.44 \times 10^{6}$ & $1.07$ & $1.37 \times 10^{8}$ \\
\hline
\nugc-H3 & $4096^3=68,719,476,736$ & 140.0  & $3.44 \times 10^{6}$ & 1.07 & $1.37 \times 10^{8}$ \\
\hline
\end{tabular}
\end{table*}

Simulations were performed by GreeM\footnote{http://www.ccs.tsukuba.ac.jp/Astro/Members/ishiya
  ma/greem} \citep{Ishiyama2009b, Ishiyama2012}, 
a massively parallel TreePM code, on the K computer at
the RIKEN Advanced Institute for Computational Science, and Aterui
supercomputer at Center for Computational Astrophysics, CfCA, of
National Astronomical Observatory of Japan.  To accelerate the
calculation of the tree force, we used the the Phantom-GRAPE
library\footnote{http://code.google.com/p/phantom-grape/}
\citep{Nitadori2006, Tanikawa2012, Tanikawa2013} with support for the
HPC-ACE architecture of the K computer and AVX instruction set
extension to the x86 architecture.

For the largest \nugc-L simulation, 
we used 16,384 nodes of the K computer. 
Each node consists of one SPARC64 VIIIfx oct-core processor
with the clock speed of 2.0 GHz and 16 GB memory. 
We performed the simulation with 16,384 MPI tasks, 
in which eight OpenMP threads ran in parallel. 
The calculation time was $50\sim60$ seconds per step 
with global and adaptive timesteps. 
The total CPU time and memory consumed were 11 million CPU hours 
(1.38 million node hours) and about 50TB. 
At $z=0$, the maximum CPU and memory imbalances are about 3\% and 25\%. 
The relatively large imbalance of memory consumption 
ensured nearly ideal load balance. 

The dataset of the particles was stored at 51 time slices from $z=20$ to $z=0$
for the \nugc-L, \nugc-M, and \nugc-S simulations.  From $z=7.54$,
total 46 output redshifts were selected, as the time interval is
proportional to the typical dynamical time of the halos.  The
corresponding logarithmic redshift interval $\Delta \log(1+z)$ is
$0.02\mbox{--}0.03$.  Moreover, six datasets at high redshifts of
$z=8.15, \ 10.0, \ 12.9, \ 16.2,$ and 20.0 were stored.  For the \nugc-H1 and
\nugc-H2 simulations, the 46 output redshifts from $z=7.54$ were
identical to those of the first three simulations. Furthermore, from
$z=12.5$ to $z=7.54$, 10 datasets with a constant logarithmic
redshift interval of $\Delta \log(1+z) = 0.02$ and two datasets at high
redshifts of $z=16.2, \ 20.0$ were stored.  Thus there are total 58 time
slices for the \nugc-H1 and \nugc-H2 simulations.  For the \nugc-H3
simulation, the output redshifts were identical to those of the
\nugc-H1 and \nugc-H2 simulations, and 22 datasets were stored down to
$z=4$.

\begin{table*}[t]
\centering
\caption{
Statistics of halos at $z=0$. 
The total number of halos is listed in the second column. 
$M_{\rm max}$ is the mass of the largest halo identified in each simulations. 
$N_{\rm max}$ is the number of particles in the largest halos, 
and $F_{\rm fof}$ is the ratio between the total mass of all halos identified 
by the FoF algorithm and that of the simulation box.}\label{tab:halo}
\begin{tabular}{lcccc}
\hline
Name & \#Halos & $M_{\rm max} (h^{-1} M_{\odot})$ & $N_{\rm max}$ & $F_{\rm fof}$ \\
\hline
\nugc-L  & $421,801,565$ & $4.11 \times 10^{15}$ & $18,685,583$ & $0.485$ \\
\nugc-M  & $52,701,925$ & $2.67 \times 10^{15}$ & $12,120,576$ & $0.485$ \\
\nugc-S  & $6,575,486$ & $1.56 \times 10^{15}$ & $7,107,526$ & $0.474$ \\
\nugc-H1 & $5,467,200$ & $4.81 \times 10^{14}$ & $17,476,256$ & $0.526$ \\
\nugc-H2 & $4,600,746$ & $4.00 \times 10^{14}$ & $116,397,797$ & $0.555$ \\
\hline
\end{tabular}
\end{table*}

\begin{figure}
 \begin{center}
  \includegraphics[width=9cm]{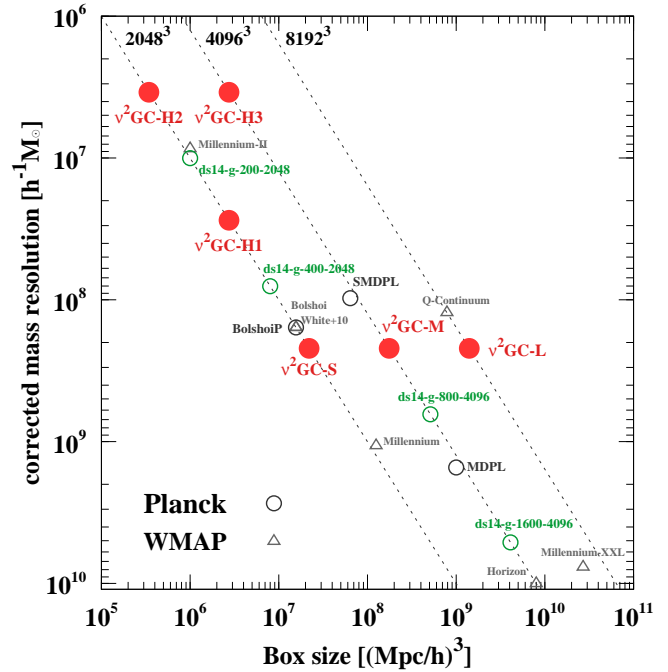} 
 \end{center}
\caption{Mass resolution versus simulation volume of recent 
large cosmological $N$-body simulations. The mass resolution of each 
simulation is corrected as the cosmological parameters of all 
simulations are the same 
as those we chose. 
The number of particles along the three dashed lines is constant.
Circles show simulations based on the Planck cosmology. 
The six red filled circles are the \nugc\ 
simulations. 
The four green circles denote four of the five Dark 
Sky Simulations (DSS; \cite{Skillman2014}). The mass resolution of the 
rest of the DSS simulations is below the range of this figure.
The three black circles are 
the BolshoiP, MDPL and SMDPL simulations \citep{Klypin2014}. 
Gray open triangles show simulations based on the WMAP cosmology
by other 
groups, Millennium simulation \citep{Springel2005}, Horizon 
\citep{Teyssier2009}, Millennium-II \citep{Boylan2009}, White+10 
\citep{White2010}, Bolshoi \citep{Klypin2011}, Millennium-XXL 
\citep{Angulo2012}, and Q Continuum \citep{Heitmann2014}.  
}
\label{fig:recent_sim}
\end{figure}

We compare the effectiveness of the \nugc\ simulations with that of the other recent large
cosmological simulations in Figure \ref{fig:recent_sim}. The figure includes
simulations with particle mass resolutions better than
$10^{10} \, h^{-1} M_{\odot}$, box sizes larger than $70 \, h^{-1} \rm Mpc$, and 
particle numbers larger than $2048^3$.  To resolve the effective
Jeans mass at high redshifts, the critical mass resolution of halos is 
$\sim 10^9 \mbox{--} 10^{10} \, h^{-1} M_{\odot}$ \citep{Nagashima2005}.
Thus, only our \nugc-L and \nugc-M simulations are able to accurately follow the physical
processes of the formation and evolution of galaxies and bright AGNs
in the context of the hierarchical structure formation scenario
based on the concordance cosmology.  
Clearly, there is a severe lack of
large simulations for small galaxies
based on the Planck cosmology, which our 
\nugc\ simulations fill.  
Just before the completion of this study, 
a simulation with volume and mass resolution comparable with those of the 
\nugc-L simulation was reported by \citet{Heitmann2014}.  However, their
simulation is still based on the WMAP7 cosmology \citep{Komatsu2011},
which is significantly different from the Planck cosmology adopted in
this study. 

Our \nugc-L simulation offers the highest mass resolution among simulations utilizing
boxes larger than $1 \, h^{-1} \rm Gpc$.
Compared with
the Millennium simulation \citep{Springel2005}, the 
\nugc-L simulation performed with 55 times more particles,
four times better mass resolution, and 11 times larger volume.
Because two times better mass resolution and three times larger volume
are obtained with the \nugc-H3 simulation, this simulation is better suited to studying
galaxies at high redshifts than the Millennium-II simulation
\citep{Boylan2009}, whose properties are comparable with those of the
\nugc-H1 and \nugc-H2 simulations.

Note that outside the regions displayed in Figure
\ref{fig:recent_sim}, there are many large simulations adopting over
 $2048^3$ particles, including Horizon Runs \citep{Kim2009, Kim2011},
MultiDark simulations \citep{Prada2012}, DEUS \citep{Alimi2012}, MICE
series \citep{Fosalba2008, Crocce2010, Fosalba2013}, Jubilee
\citep{Watson2014}, Dark Sky Simulations \citep{Skillman2014}, 
and the Outer Rim \citep{Heitmann2014}. Because
adopting a larger box in these simulations resulted in 
poorer mass resolution,
galaxy and AGN formation are not targets of these simulations 
(even in simulations with larger particles than the \nugc-L 
\citep{Skillman2014, Heitmann2014}).
Conversely, there are some studies adopted smaller boxes and higher mass
resolutions, including \citet{Ishiyama2009}, Cosmogrid
\citep{Zwart2010, Groen2011, Ishiyama2013, Rieder2013b, Rieder2013}, 
halos hosting the first stars at high redshifts
\citep{Sasaki2014}, the smallest halos first formed in the
Universe \citep{Ishiyama2014}, 
and $N$-body simulations of the Milky Way Galaxy
\citep{Bedorf2014}.
Such simulations are rare because the
simulation timestep must be small, since the dynamical time scale is
short relative to larger box simulations.  If the same number of
particles is used, simulations become less computationally challenging
as the box size increases. 

\subsection{Halo Identification}

We identified halos in each output redshift using
the Friends-of-Friends (FoF) algorithm \citep{Davis1985} with a linking
parameter of $b=0.2$. The number of particles for the smallest halo is
40. We list several halo properties across all simulations in Table
\ref{tab:halo}: the number of halos, the mass and 
number of particles in the largest halos, and the ratio between 
the total mass of all halos and that of the 
simulation box. 
The number of halos of the
largest run, \nugc-L, is 421,801,565.  
For all simulations, $\sim$50\% of dark matter exists inside halos.

The simplest definition of the halo mass is the sum of the 
masses of all member particles of an FoF group.  Yet another way to
define the halo mass is to sum the masses of the particles in a spherical region with
average density larger than the cosmic critical density by a
factor of an overdensity parameter.  A frequently used overdensity
parameter is 200 or some values according to the spherical collapse model
\citep{Bryan1998}.  In our halo catalogs, both the FoF mass and the spherical
mass with the overdensity parameter of \citet{Bryan1998} are
computed. Throughout this study, we use the FoF mass as the halo mass.

\begin{figure*}
 \begin{center}
  \includegraphics[width=16cm]{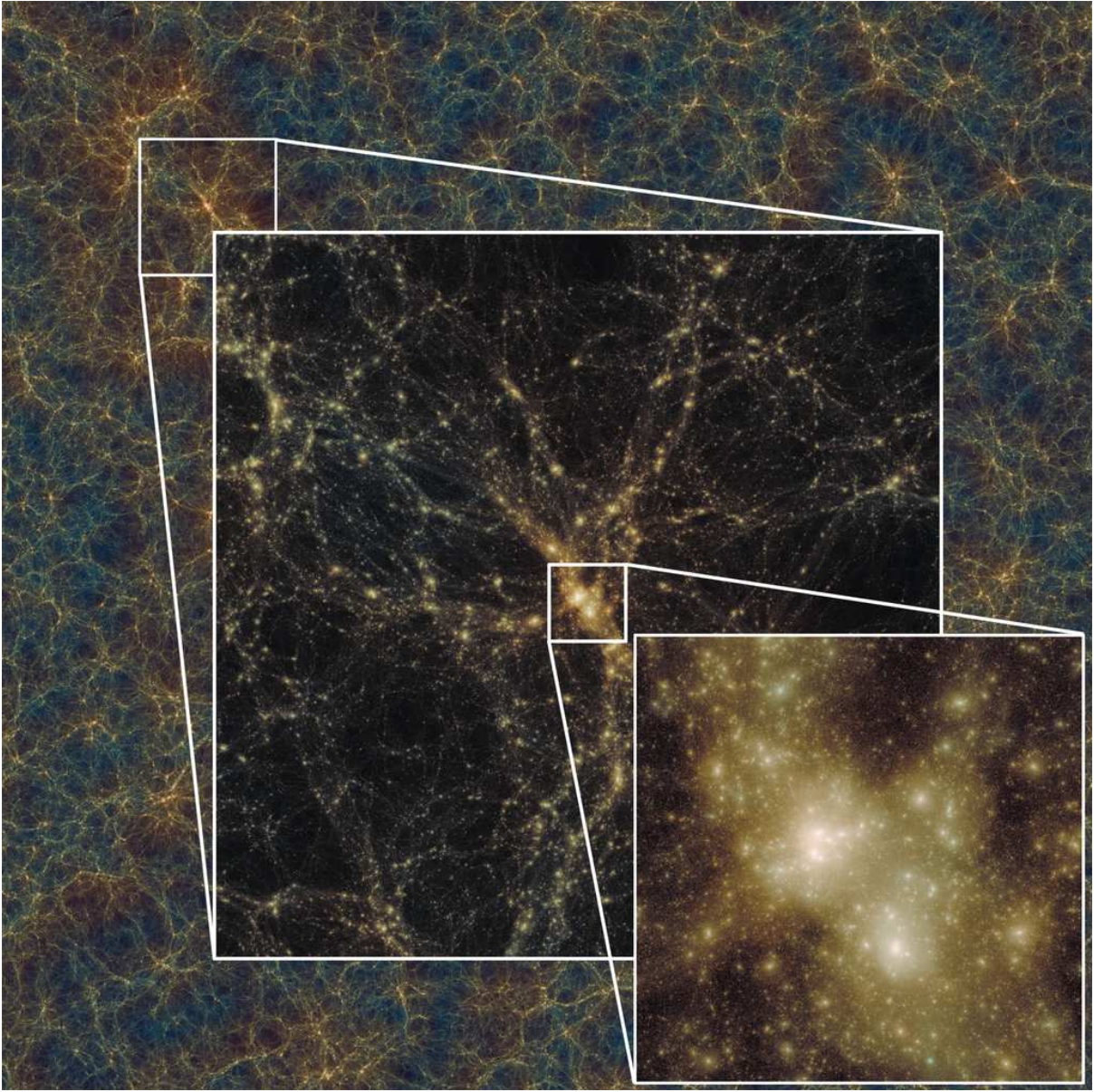} 
 \end{center}
\caption{
Dark matter distribution in the largest \nugc-L simulation at z=0.
The background image shows a projected region with a thickness of $45 \, h^{-1} \rm Mpc$
and a side length of $1120 \, h^{-1} \rm Mpc$. 
An enlargement of the largest halo is shown in the central image
with a thickness of $45 \, h^{-1} \rm Mpc$ and a side length of $140 \, h^{-1} \rm Mpc$.
The bottom right panel is a close-up of the largest halo.
}
\label{fig:snapshot}
\end{figure*}

\begin{figure*}
 \begin{center}
  \includegraphics[width=16cm]{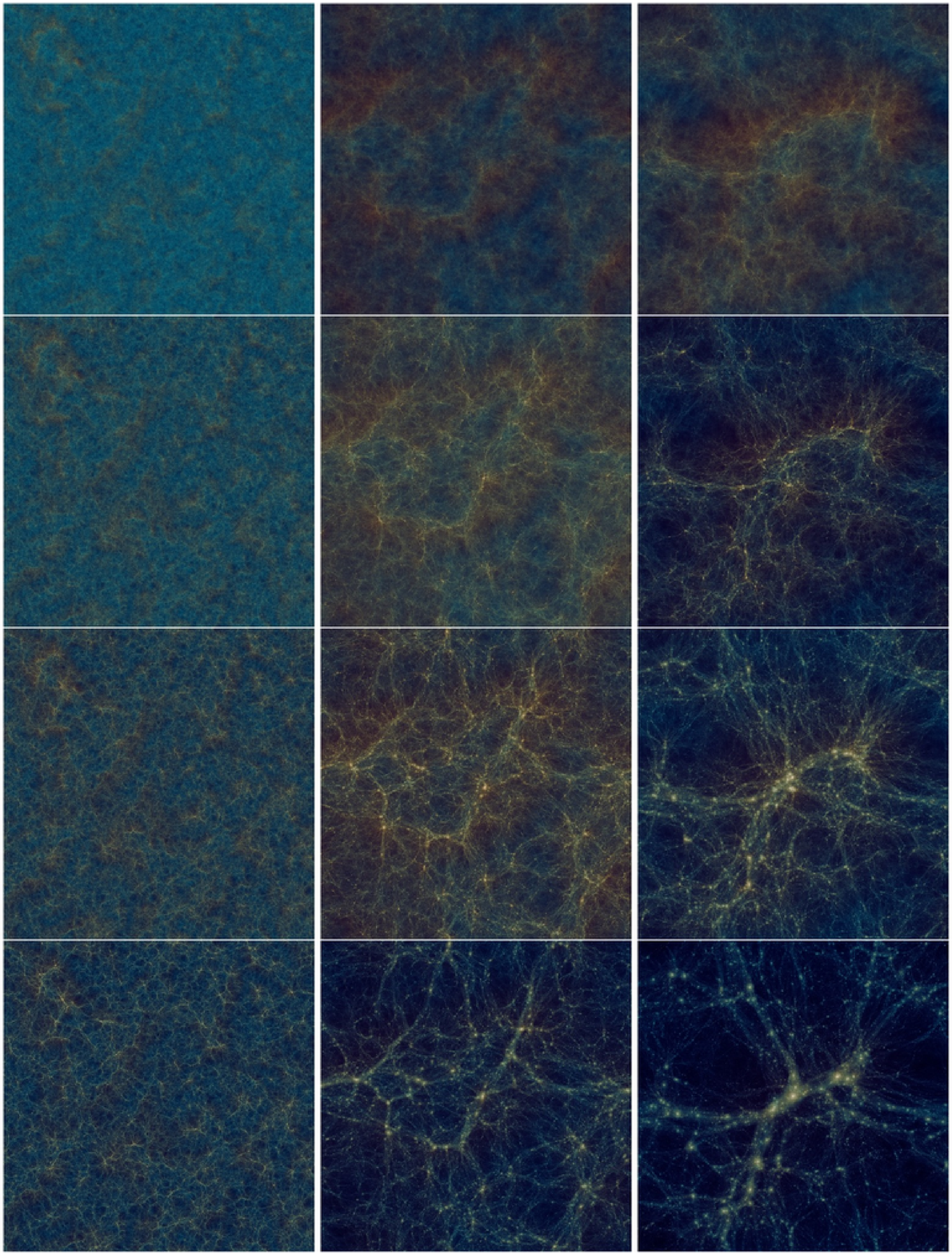} 
 \end{center}
\caption{
Time evolution of cosmic structures of the \nugc-L, \nugc-H1, and \nugc-H2 simulations 
(left to right).
From top to bottom panels, epochs at $z=7.0, \ 3.0, \ 1.0$, and 0.0 are shown. 
The thicknesses are $45 \, h^{-1} \rm Mpc$, $42 \, h^{-1} \rm Mpc$, and $21 \, h^{-1} \rm Mpc$
for the \nugc-L, \nugc-H1, and \nugc-H2 simulations, respectively.
In the images of \nugc-H1 and \nugc-H2 simulations, the largest halos are in the center.
}
\label{fig:snapshot2}
\end{figure*}

Figure \ref{fig:snapshot} shows the images of the \nugc-L simulation at 
$z=0$.  The largest halo of the \nugc-L consists of 18,685,583 particles, 
and its mass is $4.11 \times 10^{15} \, h^{-1} M_{\odot}$.  The bottom right 
panel of Figure \ref{fig:snapshot} is a close-up image of the largest 
halo. Figure \ref{fig:snapshot2} shows the redshift evolution of 
cosmic structures of the 
\nugc-L, \nugc-H1, and \nugc-H2 simulations.

\subsection{Merger Tree}

From the halo catalogs, we generated merger trees of the six simulations.
The algorithm to generate merger trees is as follows.  If a halo at
redshift $z_i$ has the largest fraction of its number of particles in common with those of
a halo at redshift $z_{i-1}$ ($z_i < z_{i-1}$), the halo at
redshift $z_i$ is assigned as the descendant of the halo at redshift
$z_{i-1}$, 
making the halo at redshift $z_{i-1}$ the progenitor of that at redshift $z_{i}$. 
A halo always has only one descendant, however, a descendant halo can 
have a number of progenitor halos. 
We used all particles belong to each FoF halo regardless of 
whether they are self-bounded. 
However, there are approaches to increase the robustness 
of merger trees by using only a subset of most bound particles 
or trajectories of halos
(e.g., \cite{Okamoto2000,Harker2006,Behroozi2013b}).

The FoF algorithm can connect physically separated halos into a single halo.
Such halos can easily fragment, emerge again at a later time, and
remerge with the host halo. Typically, the mass of such halos
is near the resolution limit. If we are not concerned with fragmentation, the
number of mergers is overestimated. Since major mergers trigger many
physical processes in the \nugc\ model, the overestimation of mergers
should be avoided.  We can infer that a large fraction of baryonic gas
in fragmented halos is lost and absorbed into the hot gas of host halos
during the first mergers because baryonic gas acts in a collisional fashion, 
i.e., in a manner different from dark matter. We forcibly removed fragmented halos
from merger trees after the first mergers.  The same algorithm was also
performed in the \ngc\ merger trees \citep{Nagashima2005}. A similar
algorithm was later called "stitch" by \citet{Fakhouri2008}.
Several algorithms to remove fragmented halos 
have been proposed 
(e.g., \cite{Helly2003, Harker2006, Fakhouri2008, Genel2009}).

The algorithm to remove fragmented halos is as follows.  If a
progenitor (redshift $z_{i-1}$) of a halo at redshift $z_i$ is also the
progenitor of smaller halos, the smaller halos are forcibly removed
from the merger tree at redshift smaller than $z_i$.  Sometimes,
progenitors of fragmented halos are not found in FoF catalogs at
redshift $z_{i-1}$ and they fragment before redshift $z_{i-2}$.  To handle
such halos, we also search the progenitors at redshift $z_{i-2}$.  If
the progenitor is also the progenitor of larger halos at redshift $z_i$,
the smaller halos is removed from the tree at redshift smaller than
$z_i$.  We performed this procedure up to redshift $z_{i-4}$.
Differently from the ``stitch'' \citep{Fakhouri2008}, the masses of
smaller halos are not added on the mass of the larger halo in our
algorithm.  Therefore, halos can lose their masses when fragmentation
of large halos occurs.

We define the most bounded particle in a halo as the marker particle of
the halo.  The spatial position of each halo (even in galaxies and AGNs)
is determined by that of its marker particle.

In our merger trees, information about subhalos is not included 
explicitly. Once halos fall into host halos and become subhalos, the 
properties of these halos after the merge, such as their masses, disappear 
in the merger tree. The evolution of subhalos in halos are modeled 
semi-analytically in the \nugc\ model. 
The spatial positions of these 
subhalos are tracked using their marker particles assigned in time slices 
when they merge into their host halos.

In the evolution of halos, physical flybys occur frequently. 
At high redshift, flybys occur with similar frequency in mergers 
\citep{Sinha2012}. 
Close flybys can influence the evolution of halos and galaxies, 
although the effect is uncertain. 
Thus, flybys are not considered in current our merger trees.

\section{Results}\label{sec:result}

In this section, we present basic quantities describing halos in the
\nugc\ simulations, halo mass function, mass accretion rate, formation
redshift, and merger statistics. We present updated fitting functions
to describe these quantities as functions of the halo mass and
redshifts. Compared with previously proposed fitting functions, we 
are able to
use unprecedentedly large simulations with higher mass and spatial
resolutions, and larger boxes (larger numbers of halos) in the 
state-of-the-art Planck cosmology.  
These differences may result in some differences in the
basic quantities.  We aim to not only extend the fitting functions for 
application to 
larger and smaller halo masses, but also provide the most advanced
fitting functions for precise cosmology and accurate models of galaxy and AGN
formation.

We exclude the \nugc-H3 simulation from the following analysis, as it
was terminated at $z=4$.  In some figures, the \nugc-M and \nugc-S
simulations are excluded because many statistics of these simulations
can not be distinguished from the largest simulation, \nugc-L.

In \S \ref{sec:result:massfunc}, 
we use $h^{-1} M_{\odot}$ as an unit of the halo mass. 
After \S \ref{sec:result:massfunc}, 
We alternatively use $M_{\odot}$ to easily compare our results
with previous studies.

\subsection{Mass Function}\label{sec:result:massfunc}

\begin{figure}
 \begin{center}
  \includegraphics[width=9cm]{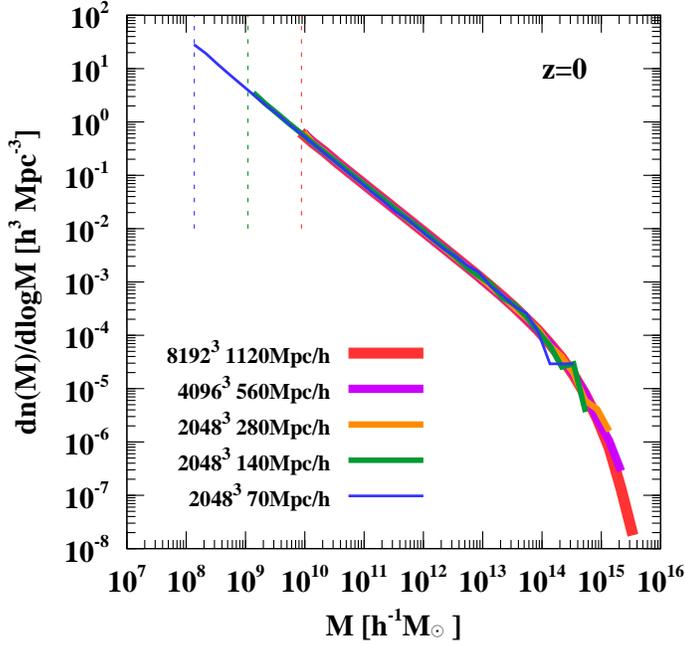} 
 \end{center}
\caption{Mass functions of dark matter halos in the \nugc\ simulations.
The vertical dashed bars show the minimum halo mass of each simulation (corresponding to 40 particles.)
}\label{fig:massfunc}
\end{figure}

In Figure \ref{fig:massfunc}, we plot the halo mass functions at $z=0$
for all \nugc\ simulations except 
for the \nugc-H3 simulation.  The five mass functions are in good agreement with 
each 
other between the mass ranges of their resolutions (40 particles, dashed
lines) and box size limits.

The halo mass functions are commonly described by 
\begin{eqnarray}
\frac{dn}{dM} = \frac{\rho_0}{M} \frac{d \ln{\sigma^{-1}}}{dM}
f(\sigma),\label{eq:massfunc}
\end{eqnarray}
where $\rho_0$ is the mean mass density, $\sigma$ is the mass
variance, and $f(\sigma)$ is an arbitrary fitting function. 
Several analytic functions \citep{Press1974, Sheth2001} 
and empirical fitting functions have been
proposed to fit well the results of cosmological simulations
(e.g., \cite{Jenkins2001, Yahagi2004, Warren2006, Tinker2008,
  Angulo2012}), and these are compiled in \citet{Murray2013}.  Here, we use
a form described in \citet{Warren2006},
\begin{eqnarray}
f(\sigma) = A \left[\left(\frac{B}{\sigma}\right)^C + 1\right]
\exp\left(-\frac{D}{\sigma^2}\right). \label{eq:massfunc2}
\end{eqnarray}
The best fit parameters to accurately describe the \nugc\ simulations 
at $z=0$ are $A=0.193, \ B=2.184,\  C=1.550$, and $D=1.186$.

\begin{figure}
 \begin{center}
  \includegraphics[width=9cm]{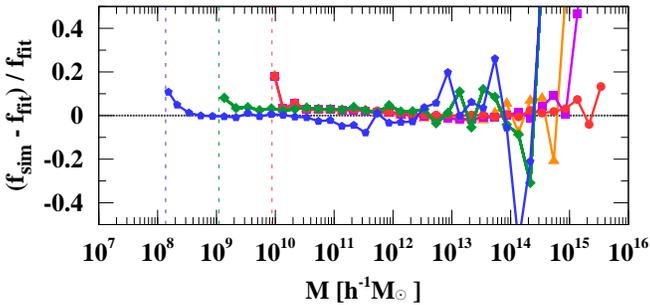} 
 \end{center}
\caption{Residuals of mass functions from the best fit function 
given by Equation (\ref{eq:massfunc2}).
The style of each curve is the same as that in Figure \ref{fig:massfunc}.
The vertical dashed bars show the minimum halo mass of each simulation (corresponding to 40 particles).
}\label{fig:massfunc_fit}
\end{figure}

\begin{figure}
 \begin{center}
  \includegraphics[width=8cm]{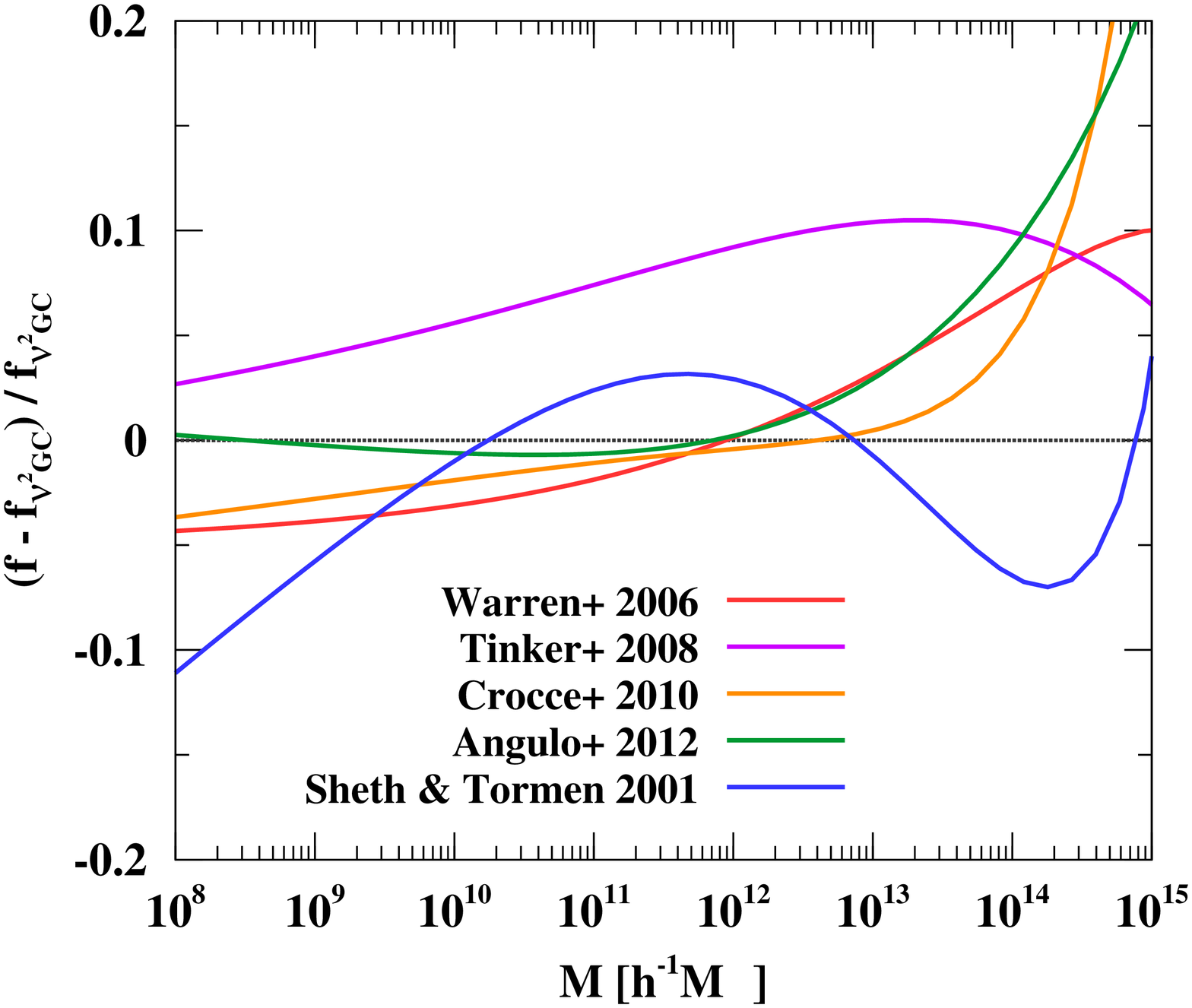} 
 \end{center}
\caption{
Comparison of our best fit mass function with functions 
proposed by other studies for redshift $z=0$. 
\citep{Sheth2001, Warren2006, Tinker2008, Crocce2010, Angulo2012}.
}\label{fig:massfunc_fit2}
\end{figure}

Figure \ref{fig:massfunc_fit} shows the residuals of the mass
functions from this best fit function.  Evidently, this function fits all
mass functions quite well for masses spanning nearly eight orders of magnitude.  
In most mass ranges, the accuracy is within 3\%.  The accuracy
worsen at very high masses because of poor statistics and at very low masses 
because of the resolution effect.

The fitting form described in Equation (\ref{eq:massfunc2}) is the
same as that used in \citet{Warren2006, Tinker2008, Crocce2010} and
\citet{Angulo2012}.  In Table \ref{tab:mfunc_pub}, we summarize the
best fit parameters of these references.  Figure
\ref{fig:massfunc_fit2} shows the residuals of these fitting functions
and the function of \citet{Sheth2001} from our best fit function.  All
functions agree well within 10\% in most mass
ranges.  For the mass larger than $10^{14} \, h^{-1} M_{\odot}$, the
difference with \citet{Crocce2010} and
\citet{Angulo2012} becomes larger.  On the other hand, the agreement
with the function of \citet{Angulo2012} is excellent for the mass
smaller than $10^{13} \, h^{-1} M_{\odot}$. 
For the mass larger than $10^{13} \, h^{-1} M_{\odot}$, 
the functions of \citet{Sheth2001, Warren2006} and \citet{Tinker2008} 
are relatively close to our fit. 

\begin{table*}[t]
\centering
\caption{
Best fit parameters of mass functions 
described in Equation (\ref{eq:massfunc2}) for redshift $z=0$.
}
\label{tab:mfunc_pub}
\begin{tabular}{lcccc}
\hline
Reference & $A$ & $B$ & $C$ & $D$ \\
\hline
This work & $0.193$ & $2.184$ & $1.550$ & $1.186$ \\
\citet{Warren2006} & $0.184$ & $2.325$ & $1.625$ & $1.198$ \\
\citet{Tinker2008} & $0.186$ & $2.57$ & $1.47$ & $1.19$ \\
\citet{Crocce2010} & $0.174$ & $2.408$ & $1.37$ & $1.036$ \\
\citet{Angulo2012} & $0.201$ & $2.08$ & $1.7$ & $1.172$ \\
\hline
\end{tabular}
\end{table*}

\begin{figure}
 \begin{center}
  \includegraphics[width=8cm]{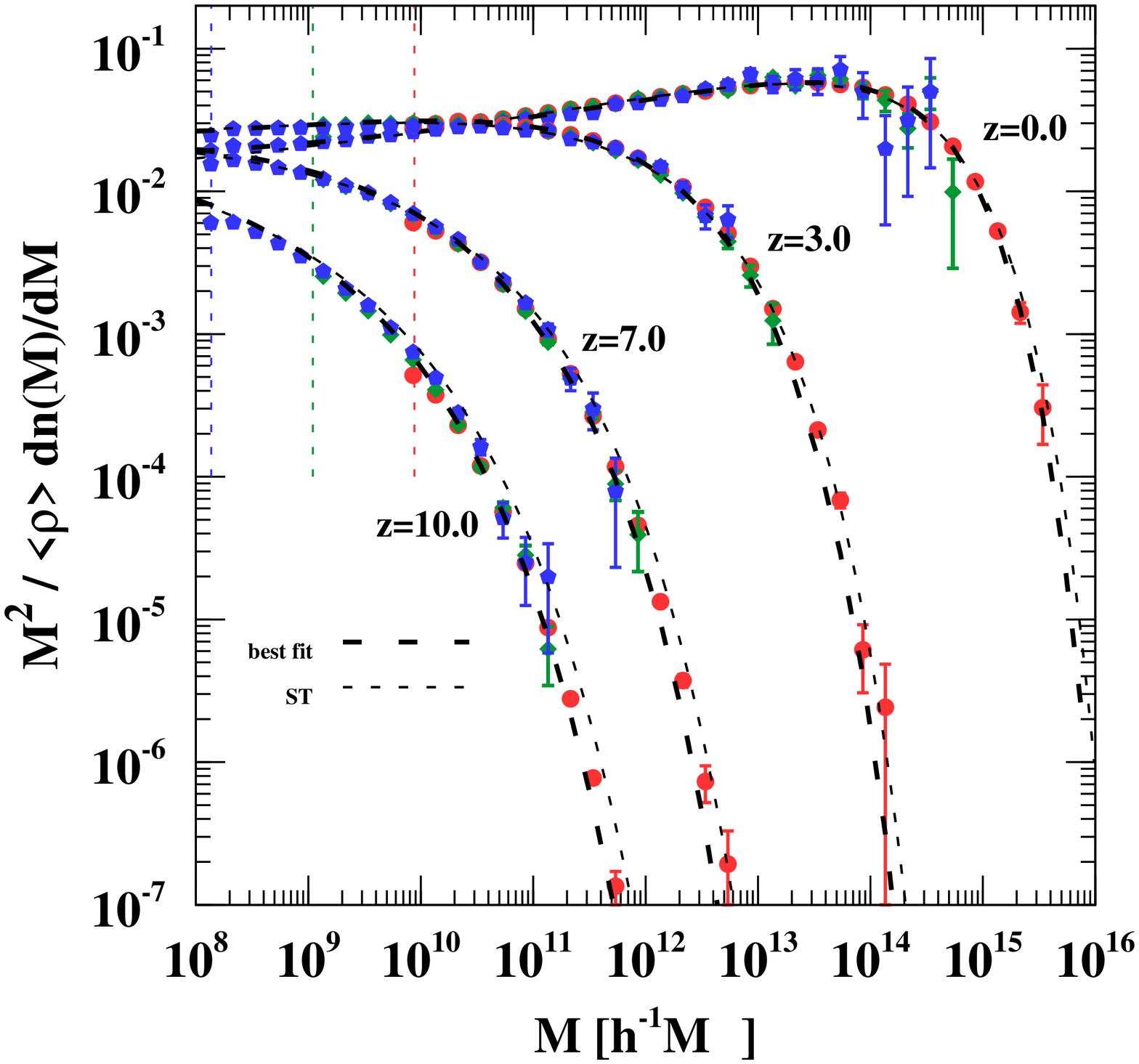} 
 \end{center}
\caption{ Multiplicity functions of the \nugc-L, \nugc-H1, and \nugc-H2
  simulations for four redshifts, $z=0.0, \ 3.0, \ 7.0$, and 10.0.  The colors of
  the symbols are the same as those used in Figure
  \ref{fig:massfunc}.  
The error bars show their Poisson error. 
The thick dashed line shows the best fit
  function given by Equation (\ref{eq:massfunc2}). 
The thin dashed line represents the function of \citet{Sheth2001}.
}\label{fig:multiplicity}
\end{figure}

In the mass function of Equations (\ref{eq:massfunc}) and
(\ref{eq:massfunc2}), the dependence on the redshift and the cosmology is
absorbed in the mass variance $\sigma$.  It is worthwhile to verify
that our best fit function can be applied to mass functions of
different redshifts.  In Figure \ref{fig:multiplicity}, we plot the
multiplicity functions of the \nugc-L, \nugc-H1, and \nugc-H2
simulations for four different redshifts, $z=0.0, \ 3.0, \ 7.0$, and $10.0$.  The
multiplicity function reduces the dynamic range in the y-direction
and helps us to understand the difference between mass functions.
Clearly, from $z=10$ to $z=0$, our best fit function could
reproduce the mass functions of the simulations quite accurately over a wide
range of masses.  Since the multiplicity functions of the \nugc-M and
\nugc-S simulations can not be distinguished from those of the
\nugc-L simulations, we did not include these functions in Figure 
\ref{fig:multiplicity} for
visualization purposes.

\begin{figure*}
 \begin{center}
  \includegraphics[width=8.4cm]{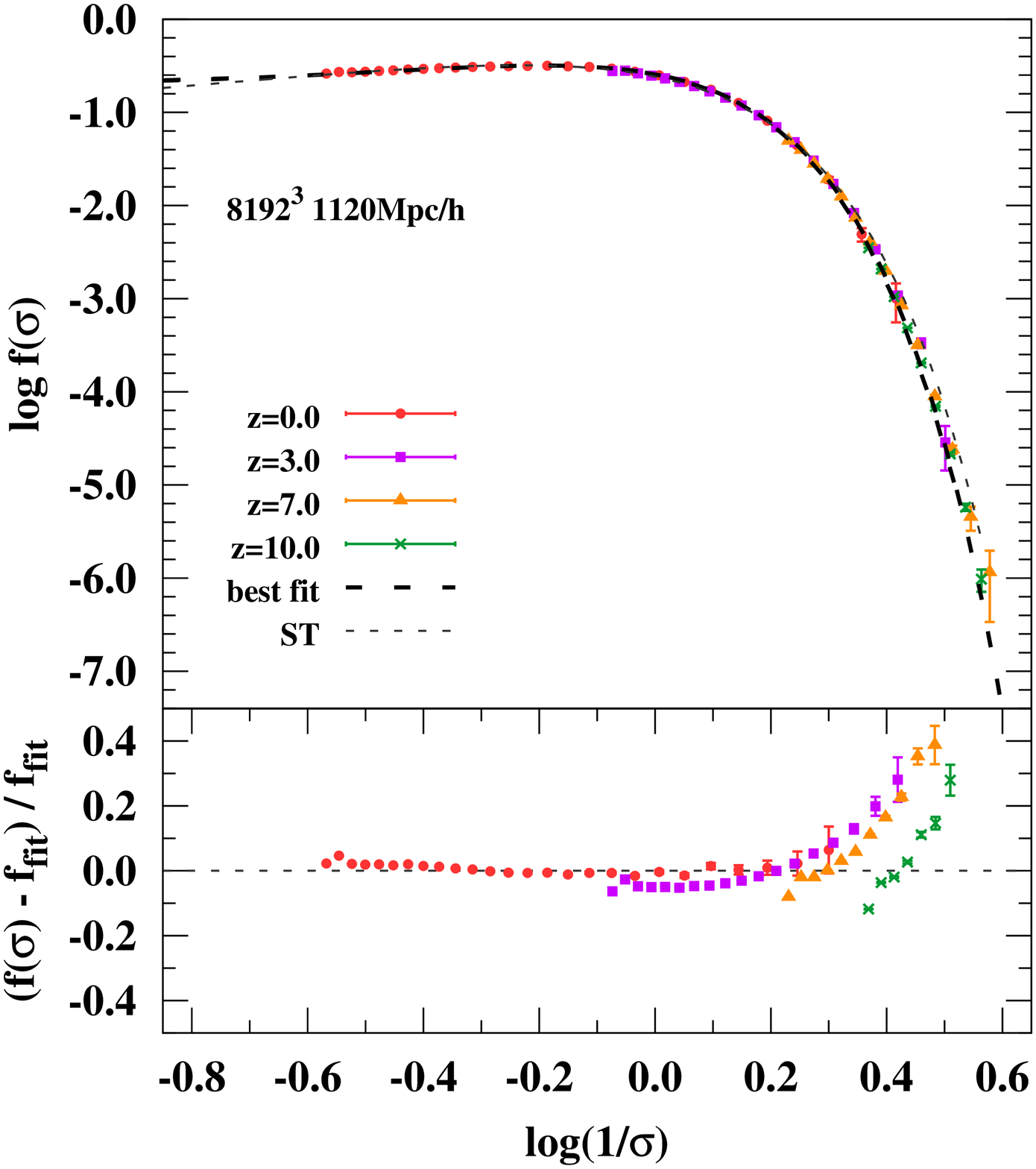} 
  \includegraphics[width=8.4cm]{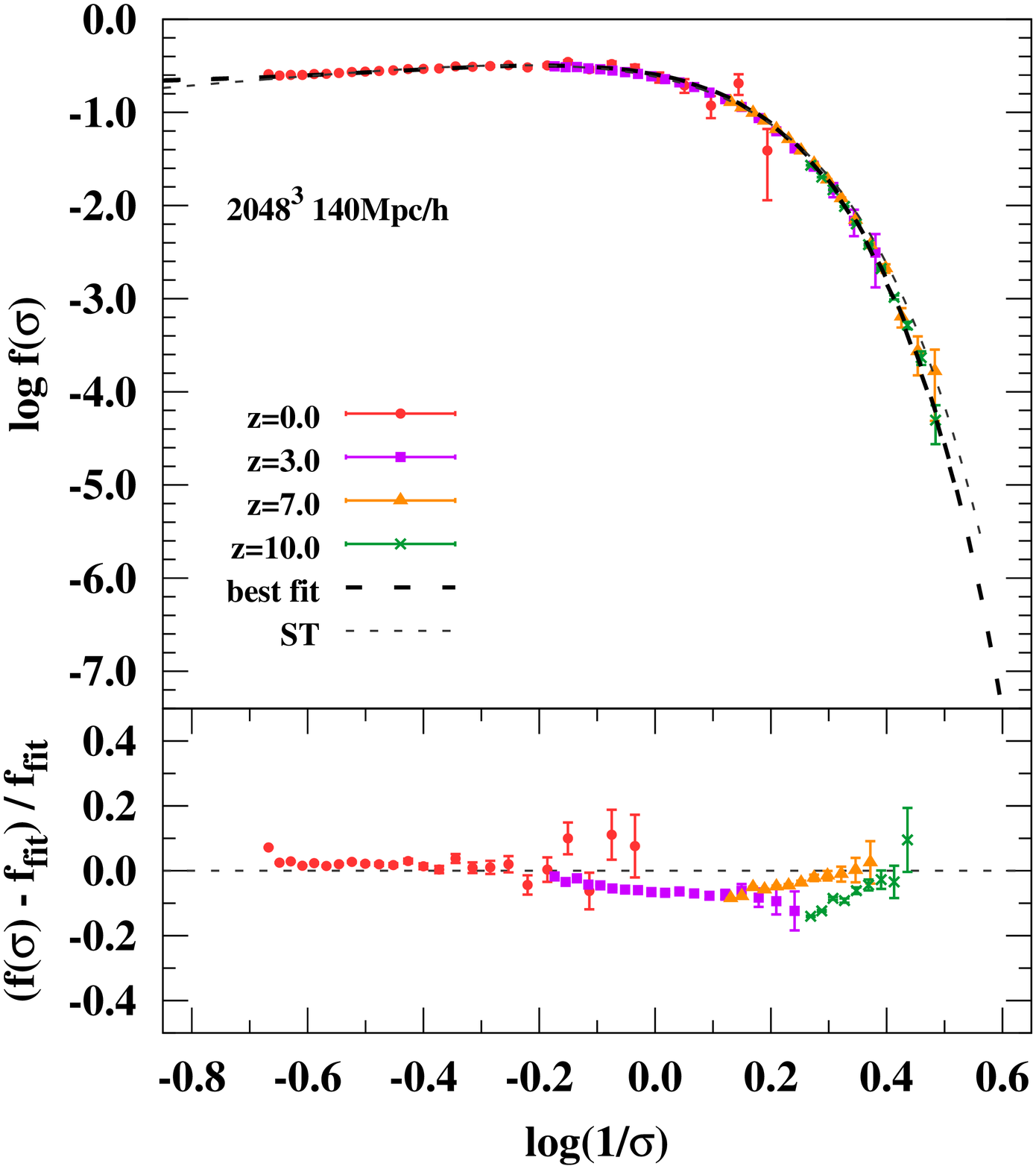} 
  \includegraphics[width=8.4cm]{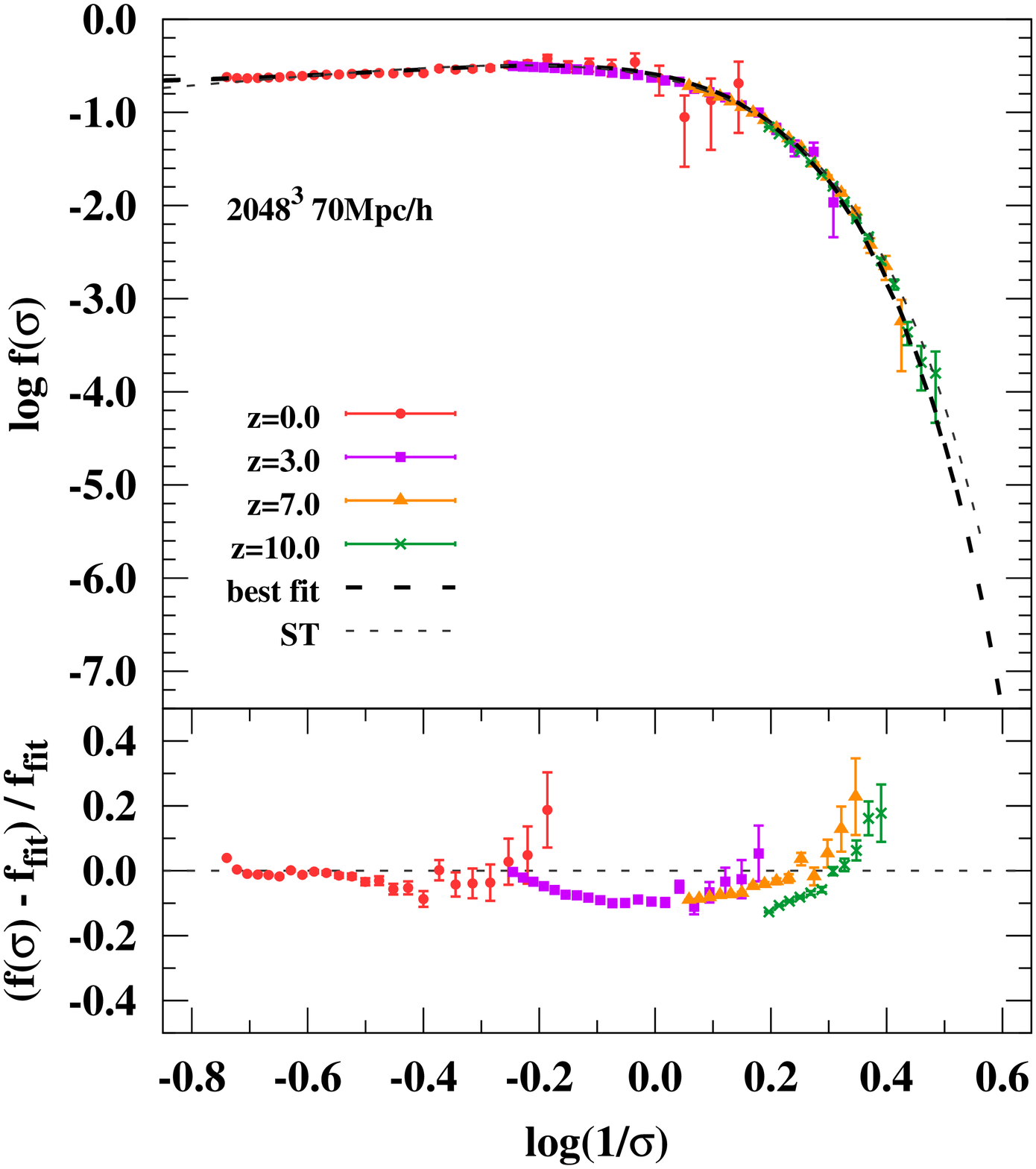} 
 \end{center}
\caption{
Another look at multiplicity functions of the \nugc-L, \nugc-H1, 
and \nugc-H2 simulations for four redshifts, $z=0.0, \ 3.0, \ 7.0$, and 10.0.
In each panel, top figure is the mass variance $\sigma$ versus $f(\sigma)$
measured directly from the simulations, 
bottom figure shows residuals from the best fit. 
The colors of the symbols are the same as those used in Figure \ref{fig:massfunc}. 
The error bars show their Poisson error. 
The thick dashed line shows the best fit function. The thin dashed line 
represents the function of \citet{Sheth2001}.
For bottom figures, only data with errors less than 10\% are shown. 
}\label{fig:massfunc_redshift}
\end{figure*}

The analytic function proposed by \citet{Sheth2001} is also plotted in Figure
\ref{fig:multiplicity}.  The difference between this function and our
best fit is negligible at $z=0$, but becomes larger as the
redshift increases.  Our fit yields slightly smaller values than 
the previously proposed one though the difference is still negligible. 
The overprediction of the model by \citet{Sheth2001} at high redshifts
is consistent with other simulations 
(e.g., \cite{Reed2007, Klypin2011, Watson2013}).

Figure \ref{fig:massfunc_redshift} gives another look at multiplicity
functions, the mass variance $\sigma$ versus $f(\sigma)$
measured directly from the \nugc-L, \nugc-H1, and \nugc-H2 simulations for four
different redshifts $z=0.0, \ 3.0, \ 7.0$, and $10.0$.  Since our best fit
function shows excellent agreement with the simulation results, excluding the
high mass region because of poor statistics, we can say that this fit
is at least universal with respect to the redshift.

In each panel of Figure \ref{fig:massfunc_redshift}, 
the residuals of $f(\sigma)$ from the best fit are also shown. 
For most areas of $-0.7 \le \log(\sigma^{-1}) \le 0.3$, 
the accuracy is within 10\%. 
Apparently, the accuracy becomes worse with increasing redshift 
for $\log(\sigma^{-1})$ larger than 0.3, simply because
our best fit is calibrated in the mass range of $5 \times 10^{8} \sim 3 \times 10^{15} \, h^{-1} M_{\odot}$ at $z=0$, 
corresponding to $-0.7 \le \log(\sigma^{-1}) \le 0.3$. 
Special attention is needed to use our best fit beyond 
$\log(\sigma^{-1})$ larger than 0.3, 
corresponding to $\sim3 \times 10^{11} \, h^{-1} M_{\odot}$ at $z=10$.

The fitting form described in Equation (\ref{eq:massfunc2}) is the
same as that used in \citet{Warren2006, Tinker2008, Crocce2010} and
  \citet{Angulo2012}, however, the best fit parameters are different, and their 
applicability to high redshifts was not confirmed.  The \nugc\ 
simulations are the first to extend this simple form to suit a wide 
range of redshifts.

\begin{figure}
 \begin{center}
  \includegraphics[width=9cm]{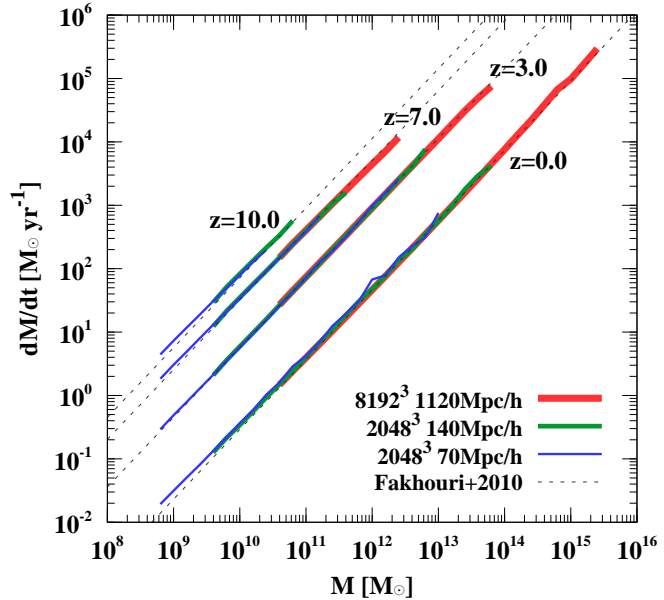} 
 \end{center}
\caption{
Mean mass accretion rate of the \nugc-L, \nugc-H1, and \nugc-H2 simulations
as a function of the halo mass
for four redshifts, $z=0.0, \ 3.0, \ 7.0$, and 10.0.
The dashed line is the relation proposed by \citet{Fakhouri2010} 
derived from the Millennium and Millennium-II simulations.}\label{fig:m-mdt}
\end{figure}

\begin{figure*}
 \begin{center}
  \includegraphics[width=14cm]{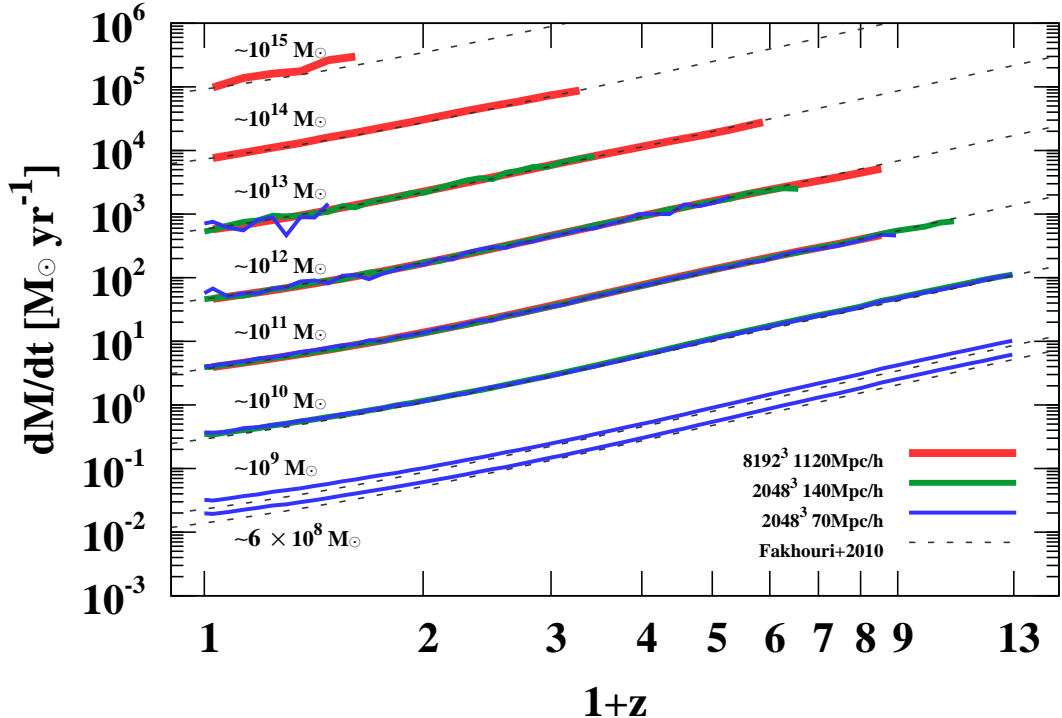} 
 \end{center}
\caption{Redshift evolution of mean mass accretion rate
of the \nugc-L, \nugc-H1, and \nugc-H2 simulations
over halo masses of approximately seven orders of magnitude.
The dashed line is the relation proposed by \citet{Fakhouri2010} 
derived from the Millennium and Millennium-II simulations.}\label{fig:z-mdt}
\end{figure*}


\subsection{Mass Accretion Rate}\label{sec:result:mar}

There are many studies for the mass accretion and assembly history
(e.g., \cite{Wechsler2002, Bosch2002, Li2007, Zhao2009, 
McBride2009, Fakhouri2010, Bosch2014, Correa2014, Correa2015}).
By tracking merger trees, we can derive the mass accretion history
of each halo as follows.  Given a halo with mass $M_i$ at redshift
$z_i$ ($i$ stands for redshift), 
we track its merger tree and identify $M_{i-1, j}$, 
the masses of its progenitors ( $M_{i-1, j} > M_{i-1, j+1}, j$ 
denotes the rank of the progenitors) at redshift $z_{i-1}$ ($z_i < z_{i-1}$).
The most massive progenitor is $M_{i-1,0}$.
We then compute the mass accretion $dM/dt = (M_i - M_{i-1, 0})
/ dt$, where $dt$ is the time interval between snapshots at redshifts
$z_i$ and $z_{i-1}$.

Hereafter, 
we use $M_{\odot}$ as an unit of the halo mass to easily compare our results
with previous studies.
Figure \ref{fig:m-mdt} shows the mean mass accretion rates of the
\nugc-L, \nugc-H1, and \nugc-H2 simulations as a function of the halo
mass.  
We only plot mass bins larger than halos of 40 particles 
and containing minimum 100 halos. 
For
$z=10$, the mean mass accretion rate of \nugc-L is not shown because
the time interval between snapshots at high redshifts is not sufficiently
small.  The three results are in good agreement with each other, reinforcing the convergence
of the \nugc\ simulations.  The dependence of the mass accretion rate on the halo mass is similar
regardless of the redshift.  We confirm that the fitting function
suggested by \citet{Fakhouri2010} (the updated fitting function of
  \citet{McBride2009} based on the Millennium simulation) agrees well
with our results (dashed line).  
This fitting function is given by \citep{Fakhouri2010}
\begin{eqnarray}
\left<\frac{dM}{dt} \right> = 46.1 \, M_{\odot} \, {\rm yr}^{-1} \left( \frac{M}{10^{12} \, M_{\odot}} \right)^{1.1} \times \nonumber \\
(1+1.11z) \sqrt{\Omega_0(1+z)^3 + \lambda_0} \ . 
\end{eqnarray}
The dependence of the mass accretion rate
per unit mass $\frac{1}{M}\frac{dM}{dt}$ on the halo mass is weak
($\propto M^{0.1}$), however, its dependence on the redshift is nearly
$\propto (1+z)^{1.5}$ at low redshifts and $\propto (1+z)^{2.5}$ for
$z>1$.

We plot the redshift evolution of the mean mass accretion rate for the 
\nugc-L, \nugc-H1, and \nugc-H2 simulations in Figure \ref{fig:z-mdt} 
over halo masses of approximately seven orders of magnitude.  Only 
redshift bins containing over 100 halos are plotted.  As shown in Figure 
\ref{fig:m-mdt}, the fitting function suggested by \citet{Fakhouri2010} 
can reproduce the dependence of the mass accretion rate on the halo 
mass and redshift fairly well.  In \citet{Fakhouri2010}, this dependence 
was confirmed in a relatively narrow mass range (from 
$10^{10} \, M_{\odot}$ to $10^{14} \, M_{\odot}$).  We found that their fitting 
function is acceptable for the Planck cosmology, and we extended their 
results to a broader mass range (from $\sim 6 \times 10^{8} \, M_{\odot}$ 
to $10^{15} \, M_{\odot}$) using the unprecedentedly high resolution and 
statistical power of the \nugc\ simulations.


\subsection{Halo Formation Redshift}\label{sec:result:zf} 
Structure formation in the Universe 
proceeds hierarchically. Smaller halos collapse earlier than larger 
halos.  We expect that a typical halo formation redshift $z_f$ decreases 
monotonically as the halo mass increases.  The extensive dataset of 
the \nugc\ simulations allows us to quantify the halo formation redshift 
over a wide mass range with strong statistics to an extent that has not 
yet been achieved.

Typically, the halo formation redshift is defined as the redshift at 
which the most massive progenitor of the main branch of a halo reaches 
its half mass compared to $z=0$.  
Fitting functions have been proposed using the 
Millennium simulations (e.g., \cite{McBride2009, Boylan2009}).  The halo 
formation redshift depends on the cosmological parameters.  In this 
section, we update previous studies for the Planck Cosmology, and 
extend them to a wider range of masses using the \nugc\ simulations.

\begin{figure*}
 \begin{center}
  \includegraphics[width=14cm]{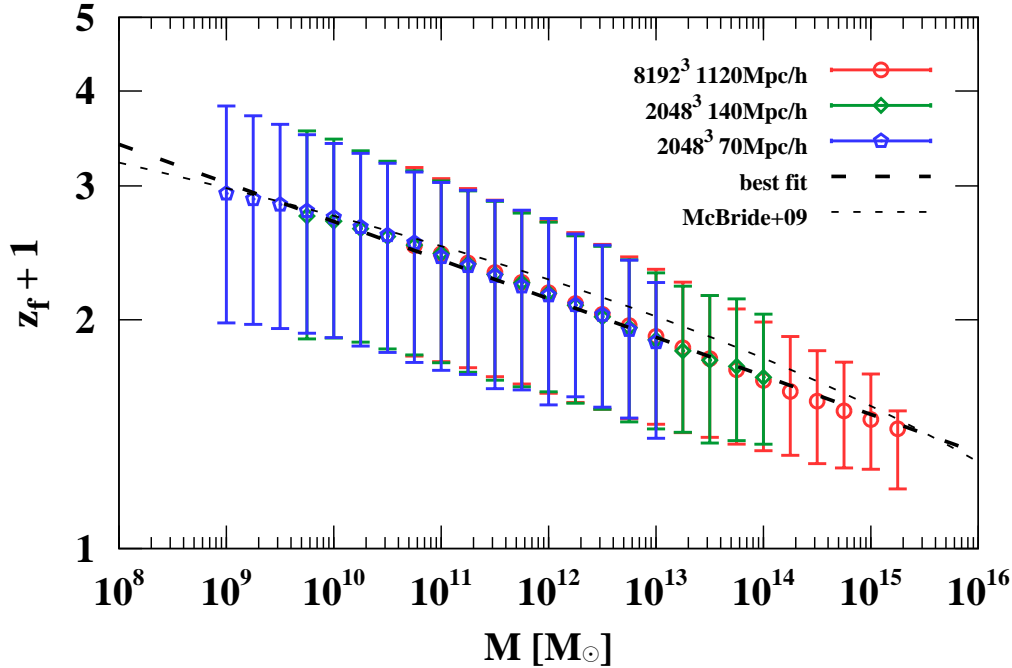} 
 \end{center}
\caption{
Mean half mass formation redshift
for the \nugc-L, \nugc-H1, and \nugc-H2 simulations
as a function of halo mass.
The error bars show the 16\% latest and earliest formation redshift.
The thick dashed line is the best fit (Equation (\ref{eq:m-zf}) in the text).
Thin dashed line is a relation proposed by \citet{McBride2009} 
derived from the Millennium simulation. 
}\label{fig:m-zf}
\end{figure*}

Figure \ref{fig:m-zf} shows the mean half mass formation redshift as a 
function of the halo mass for \nugc-L, \nugc-H1, and \nugc-H2 simulations. 
The error bars show the 16\% latest and earliest formation redshift.
The three results agree well in the range larger than their resolution limits
 (over 150 particles). The best fit function is 
\begin{eqnarray} 
1 + z_f = 2.69 \left( \frac{M}{10^{10} \, M_{\odot}} \right)^{-0.0508}. 
\label{eq:m-zf}
\end{eqnarray} 
In Figure \ref{fig:m-zf}, the fitting function obtained from the
Millennium simulations \citep{McBride2009} is plotted, and this function 
is consistent with the results but
slightly overestimates the 
average
 formation redshift for masses greater than $\sim
10^{10} \, M_{\odot}$.  The difference is $\sim11\%$ for Milky-Way-sized
halos ($\sim 10^{12} \, M_{\odot}$), 
although there are larger scatters.

\begin{figure*}
 \begin{center}
  \includegraphics[width=14cm]{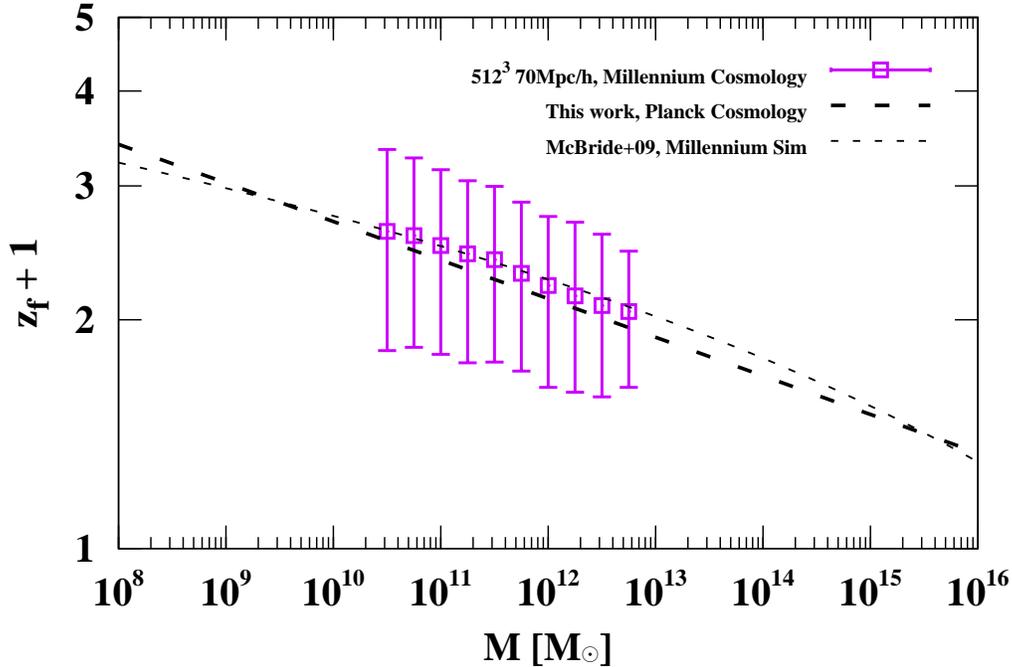} 
 \end{center}
\caption{
Mean half mass formation redshift of an additional simulation 
adopting the same cosmology of the Millennium simulations.
The error bars show the 16\% latest and earliest formation redshift.
The thick dashed line is the best fit 
derived from the \nugc\ simulations based on the Planck cosmology
(Equation (\ref{eq:m-zf}) in the text).
The thin dashed line is a relation proposed by \citet{McBride2009} 
derived from the Millennium simulation. 
}\label{fig:m-zf2}
\end{figure*}

What causes this difference? One possible cause is the
difference of methods of extracting the merger trees. Another
possibility is the difference in the adopted cosmological parameters.  The
cosmological parameters of the Millennium simulations are
$\Omega_0=0.25$, $\Omega_b=0.045$, $\lambda_0=0.75$, $h=0.73$,
$n_s=1$, and $\sigma_8=0.9$, while those of the
\nugc\ simulations are based on the state-of-the-art Planck cosmology,
namely, $\Omega_0=0.31$, $\Omega_b=0.048$, $\lambda_0=0.69$, $h=0.68$,
$n_s=0.96$, and $\sigma_8=0.83$. The striking differences between
the Planck and Millennium cosmology are in parameters $\Omega_0$, and
$\sigma_8$. Decreasing $\Omega_0$ appears to decrease the average
formation redshift. Conversely, increasing $\sigma_8$ seems to
increase the average formation redshift 
(e.g., \cite{Bosch2002, Giocoli2012}). 
Thus, the effect of the different
cosmological parameters is not trivial, as seen in the
internal structures of halos \citep{Ludlow2014, Dutton2014}.

To illustrate the cause of the difference of the mean half mass formation 
redshift between the results of the \nugc\ and Millennium simulations, 
we performed a small additional simulation with the cosmology used in 
the Millennium simulations. The number of particles used was $512^3$, 
and the comoving box size was $70 \, h^{-1} \rm Mpc$. The mass resolution was 
effectively equivalent with that adopted in the \nugc-L, \nugc-M, and 
\nugc-S simulations. In Figure \ref{fig:m-zf2}, we plot the average half 
mass formation redshifts of this simulation. Evidently, the results 
agree well with the fitting of \citet{McBride2009}. This result 
suggests that the difference in simulation results is caused not 
by the implementation of extracting merger trees but by differences in 
cosmology.

\subsection{Merger Rate}\label{sec:result:merger}

\begin{table*}[t]
\centering
\caption{
Best fit parameters for the mean merger rate 
at $z=0$ as a function of the halo mass. 
The fitting function is given by Equation (\ref{eq:merger_rate}). 
}\label{tab:merger_rate}
\begin{tabular}{ccccc}
\hline
Mass range $(M_\odot)$ & $A$ & $\beta$ & $\gamma$ & $\xi_0$ \\
\hline
$10^{15} \mbox{--} 10^{15.25}$ & $0.0741$ & $-1.917$ & $0.549$ & $0.0215$ \\
$10^{14} \mbox{--} 10^{14.25}$ & $0.0792$ & $-1.898$ & $0.472$ & $0.245$ \\
$10^{13} \mbox{--} 10^{13.25}$ & $0.0715$ & $-1.894$ & $0.583$ & $0.352$\\
$10^{12} \mbox{--} 10^{12.25}$ & $0.0357$ & $-1.965$ & $0.453$ & $0.172$  \\
$10^{11} \mbox{--} 10^{11.25}$ & $0.000788$ & $-2.340$ & $0.218$ & $0.000308$\\
$10^{10} \mbox{--} 10^{10.25}$ & $0.00159$ & $-2.360$ & $0.290$ & $0.00374$\\
\hline
\end{tabular}
\end{table*}

Given a halo with mass $M_i$ at redshift $z_i$ and its merger tree, we
computed the merger mass ratio $\xi=M_{i-1, j}/M_{i-1, 0}$, which is the
ratio between the masses of the most massive progenitor and other progenitors.
We then calculated the number of mergers per halo as a function of
$\xi$ and the descendant halo mass $M_i$.  This definition of the mean
merger rate is the same as that used in \citet{Fakhouri2008} but is different
from that discussed in \citet{Genel2009}. 

\begin{figure*}
 \begin{center}
  \includegraphics[width=14cm]{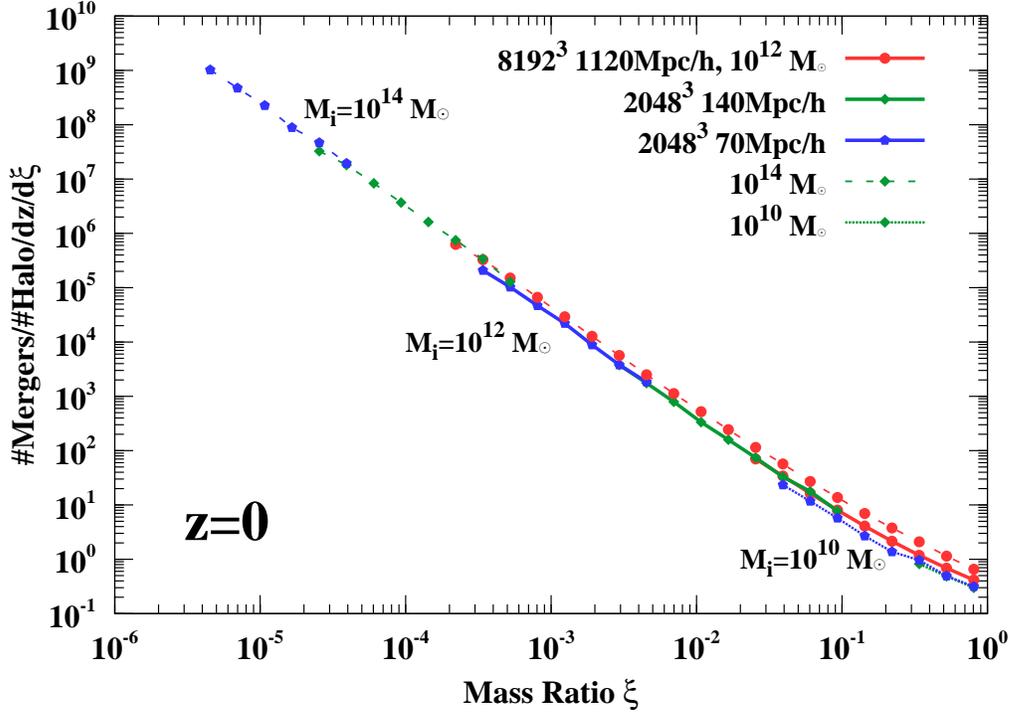} 
 \end{center}
\caption{
Mean merger rate 
of the \nugc-L, \nugc-H1, and \nugc-H2 simulations
as a function of merger mass ratio $\xi$ at $z=0$
for three descendant mass bins, 
$M_{i} = 10^{10} \, M_\odot$ (dotted), 
$10^{12} \, M_\odot$ (solid), 
and $10^{14} \, M_\odot$ (dashed curves), respectively. 
The mean merger rate is defined as the number of mergers per 
halo, $dz$, and $d\xi$. 
}\label{fig:mmr_xi}
\end{figure*}

Figure \ref{fig:mmr_xi} shows the number of mergers 
in the \nugc-L, \nugc-H1, and \nugc-H2 simulations, per halo, 
 $dz$, and $d\xi$ , as a
function of the merger mass ratio $\xi$ at $z=0$ for three 
descendant mass bins, 
$10^{10} \, M_\odot, 10^{12} \, M_\odot$, and $10^{14} \, M_\odot$.  
We only plot $\xi$ ranges in which $\xi M_{i}$ is 
larger than the halos of 100 particles. 
In addition, we exclude $\xi$ bins 
containing less than 80 mergers.
For the mean merger rate at $z=0$, we
counted the number of mergers between $z=0.027$ and $z=0.078$.  The three
simulations smoothly connect with each other although ranges with overlap
are rather narrow.  However, this configuration allows us to obtain
the relation over a wide range of the merger mass ratios $\xi$ from
$\sim 10^{-6}$ to unity.

In all mass ranges, the dependence of the mean merger rate on the mass
ratio resembles a power law starting at the smallest mass ratio ($\xi \sim 4
\times 10^{-6}$) . The slope flattens as the mass ratio increases
($\xi \ge 0.1$), which is consistent with results of previous studies \citep{Fakhouri2008,
  Genel2009, Fakhouri2010}.  The mean merger rate weakly depends on
the halo mass such that halos with higher masses have larger numbers of mergers.
Cluster-sized halos ($\sim 10^{14} \, M_\odot$) experience $\sim$1.5 more 
mergers with high mass ratio
($\xi > 0.33$) than Galaxy-sized halos ($\sim 10^{12} \, M_\odot$).

\begin{figure*}
 \begin{center}
  \includegraphics[width=14cm]{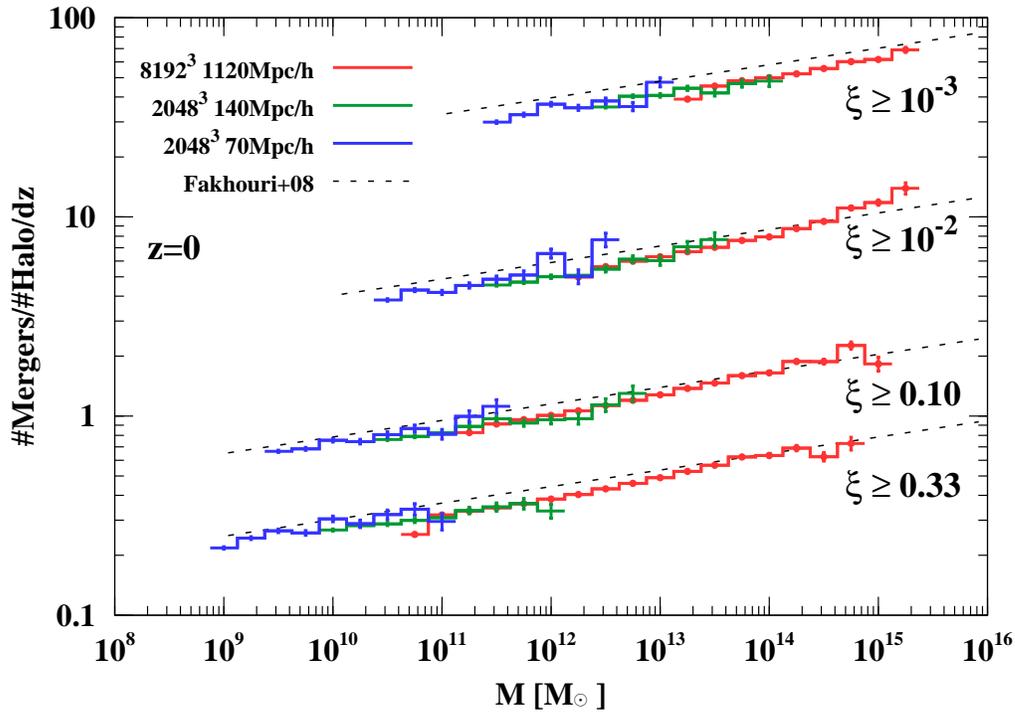} 
 \end{center}
\caption{
Integrated mean merger rate and its Poisson error
as a function of the halo mass at $z=0$
for four $\xi$ (merger mass ratio) bins, 
$\xi = 0.33, 0.10, 10^{-2}$, and $10^{-3}$, respectively. 
The dashed line is a fitting proposed by \citet{Fakhouri2008}
derived from the Millennium simulation. 
}\label{fig:mmr_z0}
\end{figure*}

The mass dependence is more striking in Figure \ref{fig:mmr_z0}, 
which shows the
halo mass dependence of the integrated mean merger rate at $z=0$ for
$\xi \ge 0.33, \ 0.10, \ 10^{-2}$, and $10^{-3}$.  Integrations were done
from $\xi$ to unity.  
The merger rate's weak dependence on the halo mass is qualitatively
consistent with previous studies (\cite{Fakhouri2008}; dashed line). 
We confirm it over a wider mass range
($10^9 \, M_\odot$ to $10^{15} \, M_\odot$) because of the high resolution and
powerful statistics of the \nugc\ simulations.
This qualitative agreement suggests 
that the mass dependence is weak and scales as $\sim M^{0.08}$ 
\citep{Fakhouri2008}.
Indeed, the number of mergers with high mass ratio
($\xi \ge 0.33$) of $10^{14.5} \, M_\odot$ halos is $\sim 2.4$
times larger than that of $10^{10} \, M_\odot$ halos. 
\citet{Fakhouri2010} updated this fitting function using 
the Millennium-II simulation. 
Because the definition of the halo mass 
in their fitting function
is different from 
that used in this study and \citet{Fakhouri2008},
we do not compare our results with \citet{Fakhouri2010}.

\begin{figure*}
 \begin{center}
  \includegraphics[width=14cm]{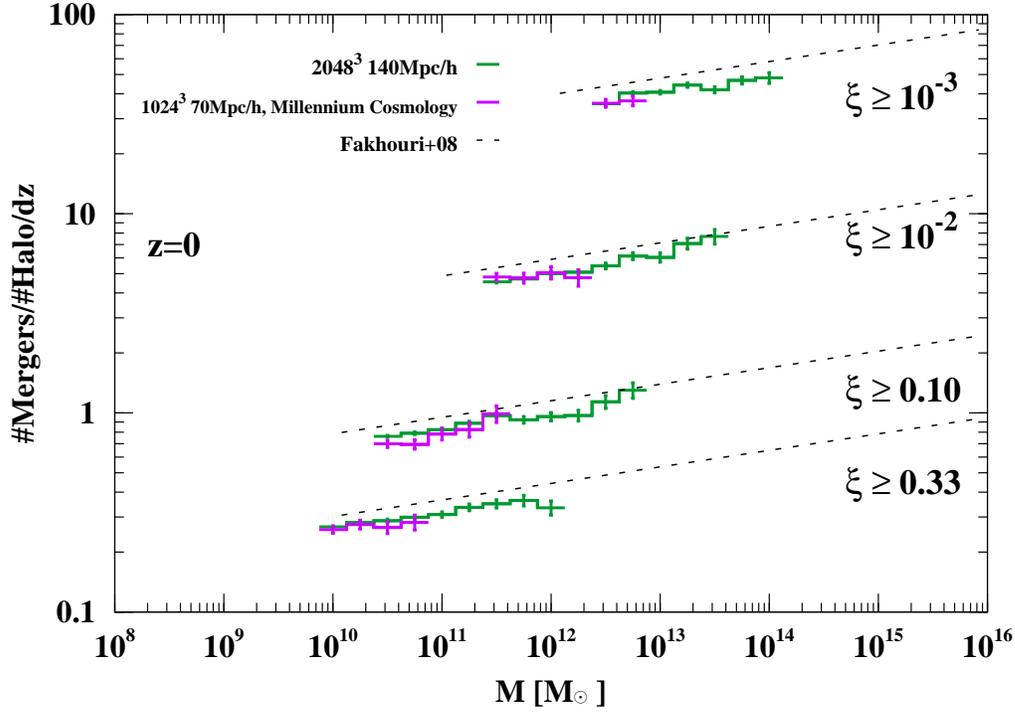} 
 \end{center}
\caption{
Integrated mean merger rate and its Poisson error
of an additional simulation 
adopting the same cosmology of the Millennium simulations at $z=0$,
for four $\xi$ (merger mass ratio) bins, 
$\xi = 0.33, 0.10, 10^{-2}$, and $10^{-3}$, respectively. 
The dashed line is a fitting proposed by \citet{Fakhouri2008}
derived from the Millennium simulation. 
}\label{fig:mmr_z0_ms}
\end{figure*}

As shown in Figure \ref{fig:mmr_z0}, 
the fitting function of \citet{Fakhouri2008} slightly overpredicts
the number of mergers, regardless of the merger mass ratio. 
To check the effect of different adopted cosmological parameters
between the \nugc\ and Millennium simulations,
we performed an additional simulation with the cosmology used in 
the Millennium simulations. The number of particles used was $1024^3$, 
and the comoving box size was $70 \, h^{-1} \rm Mpc$. The mass resolution was 
effectively equivalent with that adopted in the \nugc-H1 simulation.
In Figure \ref{fig:mmr_z0_ms}, we plot the halo mass dependence 
of the integrated mean merger rate for this and \nugc-H1 simulations. 
Clearly, the difference is tiny. 
This result implies that the difference in the 
mean merger rate might be caused by the implementation 
of extracting merger trees or the different mass resolutions.

We fit merger rates at $z=0$ in each mass bin into the following function, 
\begin{eqnarray}
\frac{1}{N_{\rm halo}} \frac{dN_{\rm m}}{d\xi dz} = A \xi^\beta \exp\left[ \left(\frac{\xi}{\xi_0}\right)^\gamma \right]. 
\label{eq:merger_rate}
\end{eqnarray}
Without terms depending on the halo mass and the redshift, this
function is the same as that used in 
\citet{Fakhouri2008} and \citet{Fakhouri2010}.  
We searched best fit parameters to suit our
simulation's results based on the Planck Universe.  The best fit
parameters are listed in Table \ref{tab:merger_rate}.  We found that
best fit parameters show large scatters, implying that the merger rate
can not be described accurately by a single universal function 
(see also \cite{Genel2009}). 
Non-universality makes it difficult to 
obtain an accurate single fitting function for the mean merger rate.

\begin{figure*}
 \begin{center}
  \includegraphics[width=14cm]{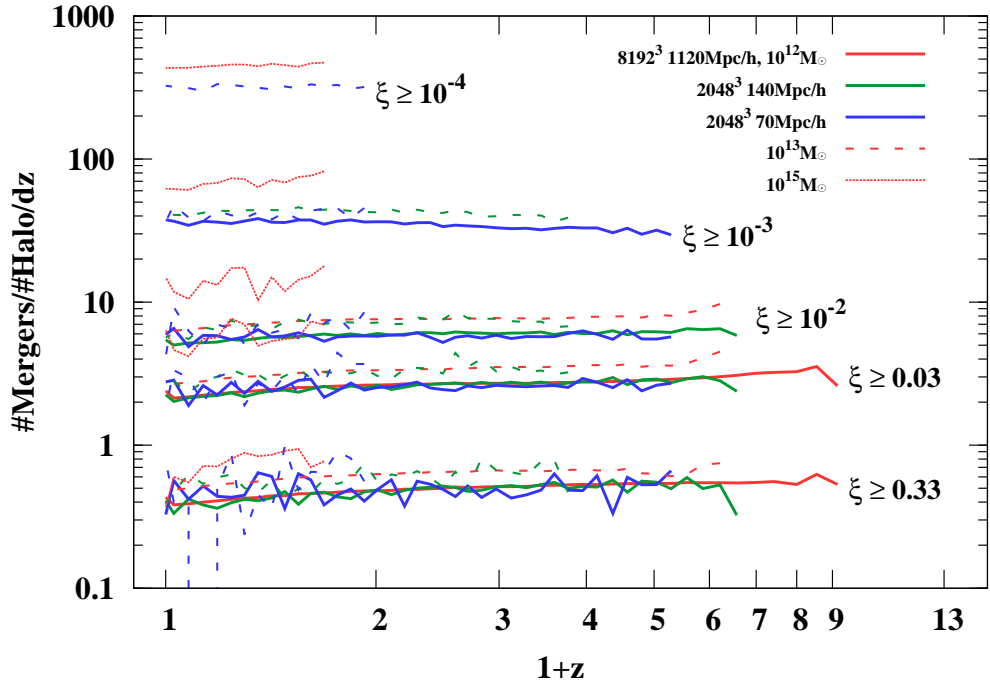} 
 \end{center}
\caption{Evolution of the integrated mean merger rate, 
for five $\xi$ (merger mass ratio) bins, 
$\xi = 0.33, 0.10, 10^{-2}, 10^{-3}$, and $10^{-4}$, 
and for three descendant mass bins, 
$M_{i} = 10^{12} \, M_\odot$ (solid),
$10^{13} \, M_\odot$ (dashed), 
and $10^{15} \, M_\odot$ (dotted curves), respectively. 
}\label{fig:mmr_evo}
\end{figure*}

The resolution of the time interval between the snapshots in
the \nugc\ simulations allow us to study the redshift evolution of the
mean merger rate up to $z \sim 6$ for the \nugc-L simulation and up
to $z \sim 10$ for the \nugc-H1 and \nugc-H2 simulations.  Figure
\ref{fig:mmr_evo} shows the redshift evolution of the integrated mean
merger rate for three different mass bins ($10^{12} \, M_\odot$,
$10^{13} \, M_\odot$, and $10^{15} \, M_\odot$) and five $\xi$ bins ($0.33$,
$0.03$, $10^{-2}$, $10^{-3}$, and $10^{-4}$).  The three simulations agree
well each other, reinforcing the convergence of the \nugc\ simulations.
The dependence on the redshift is negligible, which is consistent with
previous studies \citep{Fakhouri2008, Fakhouri2010}.  Regardless of the
redshift, a strong dependence on the merger rate and a
weak mass dependence are always observed.

\section{Discussion and Summary}\label{sec:summary}

For deeper insights regarding galaxy and AGN formation, extremely
large simulations with large volumes, large particle numbers, and high
mass resolution to resolve small galaxies are required.  We conducted
six ultralarge cosmological {\it N}-body simulations, which we called
\nugc\ simulations, on the basis of the concordance $\Lambda$CDM
cosmology consistent with observational results obtained by the Planck
satellite.  The largest simulation consists of $8192^3$ (550 billion)
dark matter particles in a box of $1.12 \, h^{-1} \rm Gpc$ (a mass resolution of
$2.20 \times 10^{8} \, h^{-1} M_{\odot}$). Among simulations utilizing
boxes larger than $1 \, h^{-1} \rm Gpc$, this is the highest resolution that has yet
been achieved in a simulation.

In this study, we presented the numerical aspects and various
properties of dark matter halos in the \nugc\ simulation suite.
Combining six large simulations, we can quantify the evolution of
halos with masses of over eight orders of magnitude, from small dwarf galaxies
to massive clusters.  With the unprecedentedly high resolution and powerful
statistics of the \nugc\ simulations, we are able to study the halo mass
function, mass accretion rate, formation redshift, and
merger statistics.  The results are summarized as follows.
\begin{enumerate}
\item 
We found that the halo mass function is well given  by 
Equation
(\ref{eq:massfunc2}), from $10^8 \, M_\odot$ to $10^{16} \, M_\odot$,
halo masses spanning nearly
eight orders of magnitude,
and from $z=10$ to $z=0$.  The analytic function of
\citet{Sheth2001} slightly overpredicts the number of halos at high
redshifts.
\item 
The halo mass accretion rate agrees well with the best fit function of
\citet{Fakhouri2010} based on the Millennium simulations.  We extended
their results to a wider mass range (from $\sim 6 \times 10^{8} \, 
M_{\odot}$ to $10^{15} \, M_{\odot}$) using the unprecedentedly high
resolution and statistical power of the \nugc\ simulations.
\item 
The half mass formation redshift is given by Equation (\ref{eq:m-zf}).
The fitting function obtained from the Millennium simulation
\citep{McBride2009} is consistent with our results, but 
slightly overestimates the formation redshift for
masses greater than $10^{10} \, M_\odot$, 
probably owing to their different adopted cosmological
parameters. 
\item 
The fitting function of \citet{Fakhouri2008}
  slightly overpredicts the number of mergers, 
  though it qualitatively reproduces the weak
  dependence on the halo mass.  The mean merger rate can not be
  described accurately by an universal function. 
\end{enumerate}

From the \nugc\ simulations, we generated mock galaxy and AGN catalogs
via our new semi-analytic galaxy formation model,
\nugc\ \makiya, which is the successor of the
\ngc\ model \citep{Nagashima2005}.  Our previous study combined a simulation 
comparable with the \nugc-S 
simulation\footnote{
The number of particles used was $2048^3$, 
and the comoving box size was $280 \, h^{-1} \rm Mpc$.
This simulation is based on the WMAP7 \citep{Komatsu2011}, namely
$\Omega_0=0.2725$, $\Omega_b=0.0455$,
$\lambda_0=0.7275$, $h=0.702$, $n_s=0.961$, and $\sigma_8=0.807$.
The merger tree was extracted by the same way described 
in this paper.}
and the \ngc\ model, successfully reproduced
the AGN downsizing trend \citep{Enoki2014}, 
and considered effects of dust attenuation on quasar luminosity functions
\citep{Shirakata2014}.
The combination of
the unprecedentedly high resolution and powerful statistics of the
\nugc\ simulations and our new model opens a new window
to study the formation and evolution of galaxies and AGNs from high to
low redshifts.  Together with Friends-of-Friends halo catalogs, the merger
trees presented in this study, and halo and subhalo catalogs generated
by the ROCKSTAR\footnote{http://code.google.com/p/rockstar/}
phase space halo/subhalo finder \citep{Behroozi2013},
we will make mock galaxy and AGN catalogs publicly available in the near future.

\bigskip
We thank the anonymous referee for his/her valuable comments.
Numerical computations were partially carried out on Aterui
supercomputer at Center for Computational Astrophysics, CfCA, of
National Astronomical Observatory of Japan, and the K computer at the
RIKEN Advanced Institute for Computational Science (Proposal numbers
hp120286, hp130026, and hp140212).  This study has been funded by 
Yamada Science Foundation, MEXT HPCI
STRATEGIC PROGRAM, MEXT/JSPS KAKENHI Grant Numbers 24740115 and 25287041.

\bibliographystyle{apj}

\end{document}